\documentclass[iop]{emulateapj}
\usepackage{epsfig, graphicx}
\usepackage{apjfonts}
\usepackage{color, rotating}
\usepackage{natbib,amsmath, amssymb}

\begin{document}
\submitted{Accepted for publication in ApJ}
\title{On the Shoulders of Giants: Properties of The Stellar Halo And The Milky Way Mass Distribution}
\author{PRAJWAL RAJ KAFLE$^{1,2}$, SANJIB SHARMA$^{1}$, GERAINT F. LEWIS$^{1}$, \& JOSS BLAND-HAWTHORN$^{1}$ }
\affil{$^{1}$Sydney Insitute for Astronomy, School of Physics A28, The University of Sydney, NSW 2006, Australia \\ 
       $^{2}$International Centre for Radio Astronomy Research (ICRAR), The University of Western Australia, 35 Stirling Highway, Crawley, WA 6009, Australia}
\email{$^{1}$ p.kafle@physics.usyd.edu.au}

\def \pasa {PASA}
\def \pasj {PASJ}
\def \jcap {JCAP}
\def \nar  {NAR}
\def \aap  {A\&A}

\def \vr {v_{r} }                   \def \vtheta {v_{\theta} }
\def \vphi {v_{\phi} }              \def \vlos {v_{\text{los}} }
\def \rsigma {\sigma_{r} }          \def \tsigma {\sigma_{\theta} }
\def \psigma {\sigma_{\phi} }       \def \sigmalos {\sigma_{\rm los} }
\def \vrot {v_{\rm rot} }           
\def \vcirc {v_{\rm circ} }           
\def \vterm {v_{\rm term} }           

\def \msun {M_{\sun} }          
\def \mvir {M_{\text{vir}}}
\def \mdisk {M_{\text{disk}}}
\def \mbulge {M_{\text{bulge}}}
\def \kms {\ {\rm kms^{-1}} }   \def \kpc {\ {\rm kpc} }  \def \feh {{\rm [Fe/H]} }
\def \pc {\ {\rm pc} }
\def \lcdm {$\Lambda \rm{CDM} \ $}

\begin{abstract}
Halo stars orbit within the potential of the Milky Way and hence their 
kinematics can be used to understand the underlying mass distribution. 
However, the inferred mass distribution depends sensitively upon assumptions made on the density 
and the velocity anisotropy profiles of the tracer population. 
Also, there is a degeneracy between 
the parameters of the halo and that of the disk or bulge. 
Most previous attempts that use halo stars have made arbitrary assumptions about these.
In this paper, we decompose the Galaxy into 3 major
components -- a bulge, a Miyamoto-Nagai disk and an NFW dark matter halo and then model 
the kinematic data of the halo Blue Horizontal Branch and K-giant stars from 
the Sloan Extension for Galactic Understanding and Exploration (SEGUE). 
Additionally, we use the gas terminal velocity curve and the Sgr A$^*$  proper motion. 
With the distance of the Sun from the centre of Galaxy $R_\odot = 8.5\kpc$,
our kinematic analysis reveals that the density of the stellar halo has a break 
at $17.2^{+1.1}_{-1.0}\kpc$, and an exponential cut-off in the outer parts 
starting at $97.7^{+15.6}_{-15.8}\kpc$.
Also, we find the tracer velocity anisotropy is radially biased with $\beta_s= 0.4\pm{0.2}$ in the outer halo.
We measure halo virial mass $\mvir$ to be $0.80^{+0.31}_{-0.16} \times 10^{12} \msun$, concentration $c$ to be $21.1^{+14.8}_{-8.3}$, 
disk mass to be $0.95^{+0.24}_{-0.30}\times10^{11}\msun$, disk scale length to be $4.9^{+0.4}_{-0.4} \kpc$ and  
bulge mass to be $0.91^{+0.31}_{-0.38} \times10^{10}\msun$. 
The mass of halo is found to be small and this has important consequences.
The giant stars reveal that the outermost halo stars have low velocity dispersion but interestingly this suggests a 
truncation of the stellar halo density rather than a small overall mass of the Galaxy.
Our estimates of local escape velocity $v_{\rm esc} = 550.9^{+32.4}_{-22.1} \kms$ and dark matter density 
$\rho^{\rm DM}_{\odot} = 0.0088^{+0.0024}_{-0.0018}\msun \pc^{-3}$($0.35^{+0.08}_{-0.07}$ GeV cm$^{-3}$) are in good agreement with recent estimates.
Some of the above estimates, in particular $\mvir$, are depended on the adopted value of $R_\odot$  
and also, on the choice of the outer power-law index of the tracer number density.
\end{abstract}

\section{Introduction}
Mass is the fundamental property of any galaxy.
An accurate measurement of the Galaxy mass has repercussions in 
many sectors, e.g., 
in its mass assembly history \citep{2002ApJ...568...52W},
identifying a realistic Galaxy in a simulation\citep[e.g.,][]{2013MNRAS.428.1696V, 2012MNRAS.424.2715W} or its analogue \citep{2012MNRAS.424.1448R}, 
simulating the tidal streams or the orbit of the satellite galaxies \citep[e.g.,][]{2010ApJ...711...32N},
studying the tidal impact of the Galaxy on the satellite galaxies 
\citep[e.g.,][]{1998ApJ...495..297J, 2013ApJ...764..161K, 2014arXiv1402.4480N} etc.
Various approaches have been undertaken to determine the mass distribution of the Galaxy,
e.g., the timing argument \citep{1959ApJ...130..705K, 2008MNRAS.384.1459L},
the local escape speed \citep{1990ApJ...353..486L, 2007MNRAS.379..755S, 2014A&A...562A..91P},
the orbital evolution of the satellite galaxies and globular clusters 
\citep{1982MNRAS.198..707L,2013ApJ...768..140B}, modeling the tidal streams
\citep{ 2005ApJ...619..807L, 2010ApJ...711...32N, 2013MNRAS.433.1826S, 2013MNRAS.435..378S},
the HI gas rotation curve \citep[e.g.][]{2009PASJ...61..227S},
fitting a parametrized model to the available observational constraints 
\citep[e.g.][]{1998MNRAS.294..429D,2011MNRAS.414.2446M,2013A&A...549A.137I} etc.
Each of this method has its own inherent limitation, for example, 
the local escape speed method suffers from the paucity of high velocity 
stars and also it is unclear whether the phase space is fully filled up to the escape velocity. 
The use of HI gas rotation curve  suffers from the fact that there is no 
extended HI disk reported for the Galaxy and hence fails to probe the mass 
that lies beyond the extent of the disk.
For an in-depth discussion and description of the various methods 
we refer the reader to the two recent reviews: 
by \cite{2013arXiv1309.3276C} on the galaxy masses
and by \cite{2013arXiv1307.8215S} on the rotation curve and references therein.

A simple, yet robust method to probe the Galaxy mass
is provided by an application of the \cite{1915MNRAS..76...70J} Equation.
The spherical Jeans Equation for a system in dynamic equilibrium is given by
\begin{equation}\label{eqn:Jeanseqn}
-\rho \frac{d\Phi}{dr} = \frac{d \rho\rsigma^2}{dr} 
                             + \rho\rsigma^2 \frac{2\beta}{r}
\end{equation}
where $\Phi$ is gravitational potential, $\rho$ is stellar density,
$\rsigma$ is radial velocity dispersion and $\beta$ is velocity anisotropy,
and all of them can be a function of galactocentric distance $r$. 
Thanks to massive stellar surveys such as SDSS/SEGUE that provide 
a catalog of position and radial velocity measurements of large number of halo tracers,
it is now possible to use the Jeans Analysis to put the most stringent constraints on the halo parameters 
out to the maximum observed distance. 
Among all the parameters that enter the Jeans Analysis, the most uncertain quantity is 
velocity anisotropy $\beta$, which is defined as
\begin{equation}\label{eqn:beta}
\beta  = 1- \frac{\tsigma^2 + \psigma^2}{2\rsigma^2}
\end{equation}
where $\tsigma$ and $\psigma$ are the velocity dispersions along the 
spherical polar ($\theta$) and azimuthal ($\phi$) directions. 
With $\beta\in[-\infty,1]$, $\beta>0$ signifies dominance of
radial motion of stars, $\beta<0$ signifies dominance of
tangential motion and $\beta=0$ an isotropic system. 
In $\beta$ is not known then the Jeans Analysis suffers from a 
degeneracy known as {\it the mass-anisotropy degeneracy}. 
In general terms it means that same radial velocity dispersion profile 
can be obtained either by lowering the $\beta$ value or by increasing the mass.
Numerous studies based on the kinematics of different stellar species, 
namely the sub dwarfs \citep{2009MNRAS.399.1223S}, the main-sequence stars \citep{2010AJ....139...59B} and 
the BHB stars \citep[][hereafter K12]{2012ApJ...761...98K} concur that $\beta \sim0.6$ (radial) in the Solar neighborhood. 
There have been recent attempts to constrain the $\beta$ beyond the Solar neighborhood, 
including K12, who use the line-of-sight velocity of a BHB sample to measure $\beta(r)$ 
to a radius of $\sim25\kpc$. 
They find a non-monotonic trend in $\beta$ starting with 0.5 (radial) 
at small $r$ which falls to -1.2 (tangential) at $r=17\kpc$ and then rises again to 0 at $\sim25\kpc$.  
An additional measurement of $\beta=0.0^{+0.20}_{-0.41}$ at $r=24\kpc$ is also
reported by \cite{2013ApJ...766...24D} in their proper motion studies of the main-sequence halo 
stars obtained from the Hubble Space Telescope (HST).    
This measurement of $\beta$ has broken {\it the mass-anisotropy degeneracy} at least out to $r=24\kpc$ (K12). 
To address the problem, practices such as reporting 
the masses for some arbitrary set of $\beta$ \citep[e.g.][]{2005MNRAS.364..433B} 
or assuming it from simulations \citep{2008ApJ...684.1143X} are a reasonable start.
However, to avoid a bias, an approach of marginalizing over all possible values of $\beta$
must be taken.
It is worth noting that an independent approach for a mass modeling using halo stars and assuming Jeans 
Equation is only a good starting point and helpful to make a testable prediction.

One preferred approach \citep{2005MNRAS.364..433B, 2008ApJ...684.1143X, 2012ApJ...761...98K} 
with the Jeans Analysis is to decompose the Galaxy into its dominant components (disk, bulge and halo). 
Inherent degeneracies among the components is a major concern of this approach
and it means assuming a higher disk and/or bulge mass would lower the halo mass and vice-versa.  
To break the degeneracies, the entire parameter space should be explored.
An alternative approach is the tracer mass formalism 
\citep{1981ApJ...244..805B,2010MNRAS.406..264W} which is
based on moments. It is a robust technique to estimate the underlying 
mass of the system provided the density and mass profiles are power laws and 
the anisotropy is constant with radius, which is
certainly not true for the Galaxy. 

In this paper, we work towards constructing a holistic model of the Galaxy
by combining best available data in hand 
such as the proper motion of SgrA$^{\text{\textasteriskcentered}}$, 
the gas terminal velocity in inner $r<R_\sun$ of the Galaxy,
and kinematics of the large number of halo tracers provided by SDSS/SEGUE.
Our aim is to provide the stellar halo number density and kinematic 
profiles out to the maximum observed distance, 
the dark matter halo mass and concentration, and the disk and bulge mass parameters.
For this we take an approach of fitting parametrized mass models of the Milky Way 
to observational constraints. 
This is largely similar to earlier works by e.g. \cite{1998MNRAS.294..429D, 2011MNRAS.414.2446M,2013A&A...549A.137I} 
but includes detailed modeling of distant halo stars.
Such a model must reproduce known local standard estimates such as 
the escape velocity $v_{\rm esc} \approx 545\kms$ \citep{2007MNRAS.379..755S, 2014A&A...562A..91P},
the dark matter density $\rho^{\rm DM}_{\odot} \approx 0.4$ GeV cm$^{-3}$ \citep{2010JCAP...08..004C,2011MNRAS.414.2446M},
the total column density $ \Sigma^{\rm total}_\odot \approx 70 \msun \pc^{-2}$ 
\citep{1989MNRAS.239..571K, 1989MNRAS.239..605K,  1989MNRAS.239..651K, 1991ApJ...367L...9K,  2013ApJ...779..115B,2013ApJ...772..108Z}  and
the angular velocity of Sun with respect to galactic center $\omega_{\odot} \approx 30 \kms \kpc^{-1}$ \citep{2004ApJ...616..872R,2010MNRAS.402..934M}.

We organized the paper as follows; 
first, in Section \ref{sec:data} we discuss the giants data, 
outline the selection criteria (for diagnostic see Appendix \ref{sec:diagnostic}) 
and estimate the distance (for full calculation see Appendix \ref{sec:distance_appendix}).
In Section \ref{sec:kinematics} we present the halo kinematical profile. 
In Section \ref{sec:model} we discuss the models for 
density, anisotropy and potential that are used to fit the kinematics of the halo. 
and in Section \ref{sec:result} we present our result and discussion.
Finally, we summarize in Section \ref{sec:conclusion}.
 
\section{DATA: GIANT STARS} \label{sec:data}
Among the wide varieties of known halo tracers, here we are interested in K giants.
These have long been studied 
\citep[e.g.][etc]{1989ApJ...339..106R, 1989ApJ...339..126R, 1990AJ....100.1191M}
to probe the distant halo. 
K giants are brighter, hence, effectively goes deeper.
Additionally, they are abundant in number in SEGUE 
\citep{2009AJ....137.4377Y}, a spectroscopic sub-survey of SDSS.
They can therefore supplement the existing catalogs of distant tracers such as the BHB stars
\citep{2000ApJ...540..825Y,2004AJ....127..899S,2008ApJ...684.1143X}
and the variable stars \citep{2009MNRAS.398.1757W, 2011ApJ...731....4S}.

\subsection{Selection Criteria}
We mine the ninth SDSS data release DR9 \citep{2012ApJS..203...21A}
to construct our giants catalog. 
The first set of selection criteria we impose to prepare the catalog are:
\begin{equation}
  \begin{cases}
    \text{Giants\ classified\ by\ SSPP} \\
    0.5 < m_g - m_r < 1.3, \\
    0.5 < m_u - m_g < 3.5, \\
    14  < m_r < 20, \\
    \log g < 2.9, \\
    \text{reddening estimate} \ E(B-V)<0.25 \ \text{and} \\    
    m_g - m_r > 0.086 \feh^2 + 0.38 \feh + 0.96, \\
  \end{cases}
\label{eqn:setA}
\end{equation}
where $m_g$ and $m_r$ are the extinction corrected magnitudes.
The metallicity [Fe/H], and the stellar parameters we use, i.e., surface-gravity $\log g$
and effective temperature $T_{\rm{eff}}$ are the one labeled in the SDSS DR9 as `ADOP',
meaning average of various estimators SSPP \footnote{SEGUE Stellar Parameter Pipeline
is a pipeline that processes the spectra SEGUE obtains} pipeline uses,
and also are the one recommended by \cite{2000ApJ...540..825Y}.
The quadratic color-metallicity cut in Equation \ref{eqn:setA}, devised by
\cite{2014ApJ...784..170X} and inferred from \cite{2008ApJS..179..326A},
is basically meant to eliminate the contamination from red horizontal branch 
and red clump stars.
Note, we take a slightly conservative cut on $\log g$ to minimize the contamination of dwarfs.
Additionally, we apply second set of selection criteria given by
\begin{equation}
  \begin{cases}
   \text{signal-to-noise ratio} \ S/N >15 \ \text{and}  \\
   \text{no critical and cautionary flags raised by SSPP}  \\
  \end{cases}
\label{eqn:setB}
\end{equation}
as a quality control cut.
This is to ensure the accuracy and the reliability of the stellar parameters 
and radial velocity values SSPP provides. 
Combination of the above given sets of selection criteria yield 5330 candidate giants. 
For the further diagnostic about selection criteria and candidate giants
see Appendix \ref{sec:diagnostic}.

As shown in next section we estimate the most probable distance of the
stars, which gives the height, $z$, of the stars from the galactic plane.
We then impose the condition $|z|>4\kpc$ to select the halo stars.
This leaves us with 5140 stars. 

\subsection{Distance Estimation}\label{sec:distance}
Correct distance measurements of the giant stars is critical for
studying  the kinematics of the halo and also for modelling the mass. 
For a correct treatment of the observational errors,
we set up the distance estimation in a Bayesian framework;
the calculations are given in full in Appendix \ref{sec:distance_appendix}.
The procedure we follow is same as given by \citet{2014ApJ...784..170X}.  
The essence of the exercise is that for each star with some 
set of observables, say $S$ = \{$m$, $c$, [Fe/H]\}, 
we obtain a corresponding absolute magnitude by matching it to color-metallicity fiducials 
of red giant sequences of clusters. 
In the end, with the inferred absolute magnitude 
and a given apparent magnitude we use standard photometric parallax 
relation to compute the distance for a star.
Instead of a single valued number, the set up allows us to 
compute a full probability distribution or posterior distribution $p(\mu|S)$ 
of the distance modulus $\mu$.

\section{The velocity dispersion profile}\label{sec:kinematics}
With the distance modulus posterior distributions $p(\mu|S)$ and the line-of-sight
velocities we now proceed to determine kinematic profiles of 
the stellar halo, namely the radial $\rsigma(r)$, polar $\tsigma(r)$ and 
azimuthal $\psigma(r)$ velocity dispersion profiles.
The line-of-sight velocities we analyze are the one provided by SSPP under the heading ``ELODIERVFINAL''.
We find that 89\% of our sample have uncertainties in radial velocity of $<10\kms$.
These velocities are in the heliocentric frame of reference and for the purpose
of modeling we need to transform them into the one centered at the Galactic center.
For this we assume the velocity of the local
standard of rest to be an IAU adopted value of $v_{\rm LSR} = 220\kms$;
the motion of the Sun with respect to the local standard of rest
to be $U_\sun=+11.1\kms, V_\sun=+12.24\kms, W_\sun=+7.25\kms$ \citep{2010MNRAS.403.1829S}
and the distance of the Sun from the center of the Galaxy to be $R_\odot=8.5\kpc$.

Had the full space motion of the stars been known, 
one could measure $\rsigma(r)$, $\tsigma(r)$ and $\psigma(r)$ by simply dividing the sample in the radial shells 
and then computing the second moment of the components of the velocity. 
Unfortunately, we only know the velocity along one direction, i.e., along the line-of-sight, 
therefore, obtaining all three dispersion profiles requires
a more careful modeling of the halo kinematics. 

Given that we are only interested in the dispersion profiles, 
we consider a gaussian velocity ellipsoid model with rotation about $z$-axis.
This model does not require any a priori knowledge of the
underlying potential or the tracer density distribution .
Generally, the velocity ellipsoid can have a tilt, although
it is evident from the recent studies 
by \cite{2009ApJ...698.1110S} and \cite{2010ApJ...716....1B}
that the tilt with respect to the spherical coordinate system for the Galactic halo
is consistent with zero so we ignore it.
Therefore, we write the velocity distribution  as a function
of radius as 
\begin{equation} \label{eqn:veDF}
p({\bf v}|\Theta,l,b,r)  = \mathcal{N}(\vr|0,\rsigma(r)) \mathcal{N}(\vtheta|0,\tsigma(r)) \mathcal{N}(\vphi|\vrot,\psigma(r)),
\end{equation}
where the model $\Theta  =\{\rsigma,\psigma,\tsigma,\vrot \}$ is given by
\begin{eqnarray}
&\rsigma(r) = {\rm Interpolation} &( \sigma_{r, m} , r_{m} )      \nonumber \\
&\tsigma(r) = {\rm Interpolation} &( \sigma_{\theta, m} , r_{m} ) \nonumber \\
&\psigma(r) = {\rm Interpolation} &( \sigma_{\phi, m} , r_{m} )   \nonumber, \text{and} 
\end{eqnarray}
the notation $\mathcal{N}$ represents the gaussian distribution
centered at $\bar{x}$ and with dispersion $\sigma$ given by
\begin{equation} \label{eqn:gaussian_dist}
\mathcal{N}(x|\bar{x},\sigma) = \frac{1}{\sqrt{2\pi} \sigma} {\rm exp}
 \left[ -\frac{(x - \bar{x})^2}{2\sigma^2} \right].
\end{equation}
Here, $r_m$ are grid points in radius $r$, which we call nodes.
Each of these nodes will have a corresponding value of the velocity dispersion.
Thus the final dispersion profile is obtained from a linear
interpolation over the nodes. 

While the location of the nodes can be fixed arbitrarily,
for a more systematic approach, we choose them such that 
for $r<70\kpc$ each node has 500 stars. 
For $r>70\kpc$, due to fewer stars, we choose them such that
each node has 30 stars.
This is a non-parametric approach to obtain a kinematic profile
and is a useful technique in our case for two reasons. 
First, we do not have an exact distance but
a probability distribution of distance modulus $p(\mu)$ of each star in our sample.
Hence, unlike previous studies, e.g., \cite{2012ApJ...761...98K, 2013MNRAS.430.2973K}
the data cannot be segregated into the radial bins because 
a star near a bin edge could have some finite probability to be in a neighboring bin.
In fact, depending on the distance probability of each star, 
this approach enables each star to make appropriate 
contribution to each node where $p(\mu)$ is non zero.
Second, in \cite{2012ApJ...761...98K} undulations were reported in 
the kinematic profiles of the BHB stars. Hence, we do not want to restrict our analysis 
by making an assumption about the functional forms
of $\rsigma$, $\tsigma$, and $\psigma$. 

In the absence of proper motion information, we marginalize Equation \ref{eqn:veDF} over 
the tangential velocities, $v_l$ and $v_b$.
The resultant marginalized distribution function (DF) can be expressed as,
\begin{equation}\label{eqn:vlosd}
p( \vlos| \Theta, l, b) \propto \iiint p( {\bf v}| \Theta ,l,b,\mu) p(\mu) \ d v_l \ d v_b \ d \mu.
\end{equation}
Above DF is convolved with distance modulus posterior of each star 
$p(\mu)$ from Equation \ref{eqn:bayes}.
The convolution corrects for the spatial selection effect and additionally, also 
enables to propagate the distance uncertainties to our final estimate of the kinematic profiles. 

Finally, to obtain the dispersion profiles we use the likelihood 
estimation technique based on Markov Chain Monte Carlo (MCMC) random walks.
For all experiments below, we use a MCMC algorithm, namely a stretch move as
described in \cite{Goodman} to sample from the posterior probability distribution given by our model. 
Our MCMC walks were run for sufficient autocorrelation time, so as to ensure that 
the distributions of parameters were stabilized around certain values. 
Advantage of this method over standard MCMC algorithms such as Metropolis-Hastings
is that this method explores the parameter space efficiently and also, 
produces independent samples with a much shorter autocorrelation time.

The log-likelihood function, $\mathcal{L}$, we use is 
\begin{equation} \label{eqn:likelihood}
\mathcal{L} (\Theta) = \sum_{i=1}^{n} \log p(v_{\text{los},i}|\Theta,l_i, b_i), 
\end{equation}
where sum is over the total number of stars $n$. 
The MCMC run gives us the posterior distributions of the 
model parameters $\Theta=\rsigma, \tsigma, \psigma, \vrot$ at given distances.
The values corresponding to the highest likelihood are considered as the best 
estimates of the model parameters and the uncertainties are computed from 
the $16^{\rm th}$ and $84^{\rm th}$ percentile of the distributions.

\begin{figure*}
  \centering
   \includegraphics[width=0.8\textwidth]{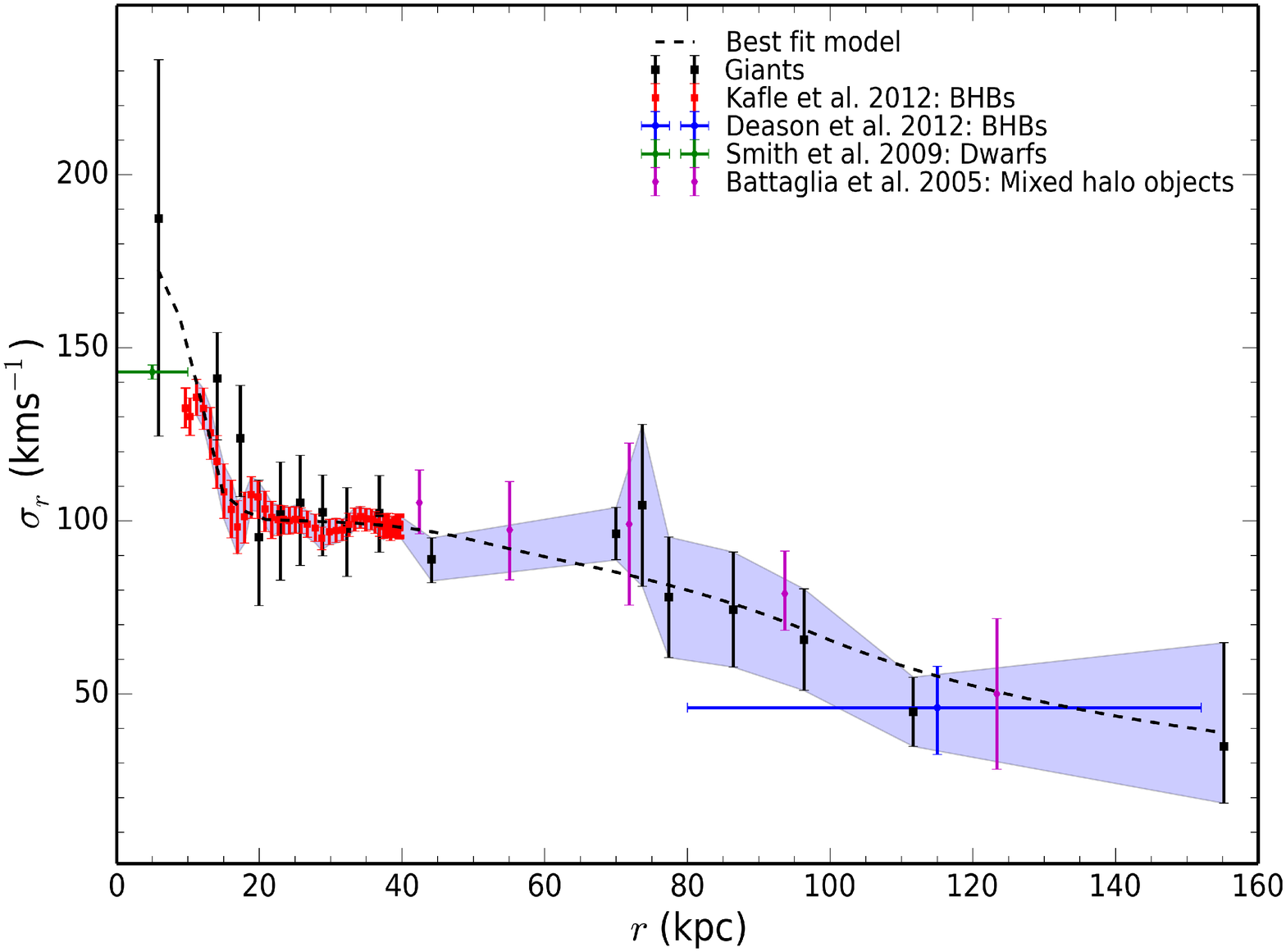}
      \caption{Radial velocity dispersion profile, $\rsigma(r)$, of the stellar halo:
             the black squares with error bars are the $\rsigma$
             values for the giants computed in this paper, the red squares with error bars are the                        
             estimates for BHB stars taken from K12,
             the blue dot with error bar is the measured value for BHB stars from \cite{2012MNRAS.425.2840D},
             the magenta dots with error bars are \cite{2005MNRAS.364..433B} combined estimate for 
             mixed sample of halo objects populating at large $r$ 
             and the green dot with horizontal and vertical error bars
             are the reported value for SEGUE sub-dwarfs by \cite{2009MNRAS.399.1223S}.
    	     The dashed black line is our best fit model for a combined BHB and giants sample 
             (the shaded purple region).}
\label{fig:rsigma}
\end{figure*}

Our $\rsigma$ profile of the halo giants is shown in Figure \ref{fig:rsigma}
by black squares. 
It starts high at $190\kms$, in the inner region, 
drops to $100\kms$ at distance $r\sim20\kpc$ and then remains flat till $r\sim70\kpc$.
This is consistent with the trend seen in BHB stars \citep{2012ApJ...761...98K},
shown by red squares.
However, note the break in $\rsigma(r)$ profile of BHB stars
is at a radius of $\sim17\pc$,
which is slightly smaller than that of the giants.
We suspect this to be due to larger distance errors of
giants. Distance errors will have the effect of smoothing sharp transitions.
At radius $r>70 \kpc$, for giants we find that there is a further drop in the $\rsigma$
reaching as low as $35\kms$ at $\sim155\kpc$.
The magenta data point taken from \cite{2005MNRAS.364..433B} also shows a low $\rsigma$ 
at such a large distance. 
Similarly, in the range $r\sim100-150\kpc$ \cite{2012MNRAS.425.2840D}
find a low $\rsigma \approx50-60\kms$ among their BHB sample
(blue data).
Similar trends in $\rsigma(r)$ for different populations reported above 
is reassuring, since they trace the same gravitational potential.
The large error bar in $\rsigma(r)$ value for the giants sample at $r\sim5.9\kpc$ is 
because, close to the Sun and along the galactic pole, the 
contribution of the radial velocity to the line-of-sight velocity is low.

\begin{figure}
  \centering
    \includegraphics[width=0.48\textwidth]{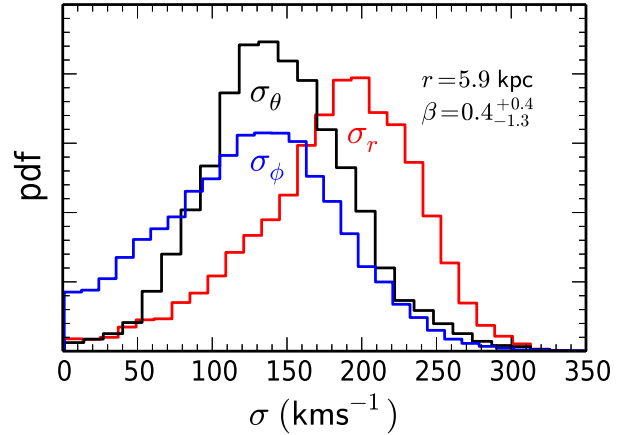}
    \caption{Posterior distributions of the velocity dispersions of the giants at $r=5.9$ kpc:
             the red, black and blue lines show the probability distribution functions
             for $\rsigma, \tsigma, \psigma$ respectively.}
\label{fig:beta5.88}
\end{figure}

Although we can measure $\rsigma$, we are unable to constrain  
the tangential components of the dispersion, $\tsigma$ and $\psigma$, for the following reasons.
First, with a line-of-sight component of the velocity alone we cannot measure the 
tangential dispersions at distances $r \gg R_\sun$.
Second, the distance uncertainty of our sample of giants is large, rendering 
large uncertainties in the tangential dispersions.
However, at the first node $r\sim5.9$ kpc the tangential velocity
  contributes significantly to $v_{\rm los}$, which makes it
possible to compute $\tsigma$ and $\psigma$, see Figure \ref{fig:beta5.88}. 
Here we find that the halo is radial, $\beta = 0.4^{+0.4}_{-1.3}$,
which is in agreement with the previous results 
using the subdwarfs \citep{2009MNRAS.399.1223S}, 
the main-sequence stars \citep{2010ApJ...716....1B} and the BHB stars 
\citep{2012ApJ...761...98K} at a similar distance.  

The issue we ignore in this study is the effect of
substructures on kinematics. In K12, it was shown that the
effect of the two most dominant structures in the halo, 
the Sagittarius stellar stream and the Virgo over-density,
is negligible. 

\section{Fitting the model}\label{sec:model}
We now proceed to modeling the $\rsigma(r)$ profile of the halo
in order to probe the Galactic potential, $\Phi(r)$. 
For this we rearrange the Jeans equation (\ref{eqn:Jeanseqn}) as 
\begin{equation}\label{eqn:Jeanseq_rsig}
\frac{d\rsigma(r)^2}{dr} = -\frac{1}{r} \left[ r \frac{d\Phi(r)}{dr} + \rsigma(r)^2 
                            \left( 2 \beta(r) + \frac{d\ln \rho(r)}{d\ln r} \right) \right]. 
\end{equation}
We solve this first order differential equation in $r$ numerically, 
forcing a boundary condition $r\to\infty$ as $\rsigma \to0$. 
For some supplied profile for potential $\Phi$, density
$\rho$ and anisotropy $\beta$,  
we use the numerical solution for $\rsigma(r)$ to fit the observed $\rsigma(r)$ obtained in 
Section \ref{sec:kinematics} (Figure \ref{fig:rsigma}). 
In the following sections we describe parametric forms of $\Phi(r)$, $\rho(r)$, and $\beta(r)$ 
that are used in our model, for a quick reference see Table \ref{table:models}. 

\begin{table*}
\normalsize
\caption{Model prescription of the assumed components of the Galaxy.}
\centering
\begin{tabular}{ c  p{3.cm} c  p{4.9cm} }
\hline\hline 
&                   &               &         \\
& Physical Quantity & Model & Comment \\ \cline{2-4}
&                   &               &         \\                      
\rotatebox[origin=r]{90}{Stellar Halo} & Density & 
$\rho(r) \propto 
        \begin{cases}
	  (r/r_b)^{-2.4} & \text{if } r<r_b \\
          (r/r_b)^{-4.5} & \text{if } r_b \leqslant r < r_t \\
          (r_t/r_b)^{-4.5} \ (r/r_t)^{\epsilon} \ {\rm exp} \left[ -\frac{r-r_t}{\Delta} \right] & \text{if } r \geqslant r_t.
        \end{cases}$
 & 
\vspace{-0.6cm} $r_b$ is the break radius, $r_t$ is the truncation radius and $\Delta$ is the scale length of fall.
  Assuming, the logarithmic slope at $r=r_t$ is continuous gives 
  $\epsilon = \frac{r_t}{\Delta}-4.5$ \vspace{0.4cm}\\ 
 & Anisotropy & 
$\beta(r) = 
        \begin{cases}
	 \text{Interpolate(Observed values)}   & \text{if } r\leqslant25\kpc \\

	\begin{cases}
	  \text{Max}\{\frac{r - 25}{r_2 - 25} \ \beta_{\rm s} ,
	  	          \beta_s \} & \text{if } \beta_s<0 \\ 
	  \text{Min}\{\frac{r - 25}{r_2 - 25} \ \beta_{\rm s} ,
                   \beta_s \} & \text{if } \beta_s\geqslant0 \\
	  \end{cases}
			  & \text{if } r > 25\kpc
         \end{cases}$
& 
\vspace{-0.8cm} Observed values are taken from the literature 
\citep{2012ApJ...761...98K, 2013ApJ...766...24D},
25 is maximum observed distance in kpc, 
$\beta_s$ is the velocity anisotropy outside radius $r_2$. \\ &&&\\
\rotatebox[origin=c]{90}{Dark Halo} & NFW potential \citep{1996ApJ...462..563N}& 
$\begin{aligned}
        \Phi_{\rm NFW} = &\frac{- \text{G} M_{\rm vir} \ln(1+rc/R_{\rm vir})}{g(c)r}, \\
        g(c) = &\ln(1+c) - \frac{c}{1+c}, \text{and} \\
        R_{\rm vir} = &\left( \frac{2M_{\rm vir} \text{G}}{H_0^2 \Omega_m \delta_{\rm th}} \right)^{1/3}\\
 \end{aligned}$
& 
\vspace{-1.2cm} $\mvir$, $c$ and $R_{\rm vir}$ are the virial mass, concentration and the virial radius respectively.
$\delta_{\rm th}$ is an over-density of the dark matter compared to the average matter density, 
$H_0$ is the Hubble constant and $\Omega_m$ is the matter density of the universe. \\ 
&&&\\
\rotatebox[origin=C]{90}{Disk} & \cite{1975PASJ...27..533M} potential &
$\begin{aligned}
\Phi_{\rm disk}(R,z) =  - \frac{G \mdisk }{\sqrt{R^2 + (a+ \sqrt{z^2 + b^2})^2}}\\
\end{aligned}$
& $a$, $b$ and $\mdisk$ are the scale length, scale height and mass respectively.  \label{row:disk_pot} \\ 
&&&\\
\rotatebox[origin=c]{90}{Bulge} 
& Spheroidal bulge \citep[Equation 2.207 a and b]{2008gady.book.....B} density & 
$\begin{aligned}
      \rho_{\rm bulge}(R,z) &=  \rho_{b0} \left( \frac{m}{a_b} \right) ^{-\alpha_b} {\rm e}^{-m^2/s^2}, \text{where}\\
      m &= \sqrt{R^2 + z^2/q_b^2} 
\end{aligned}$
& $\rho_{b0}$ is a density normalization, which is a function of $M_{\rm bulge}$. $s$ is a scale length. 
Values for remaining parameters $a_b$, $\alpha_b$ and $q_b$ are assumed to be 
1 \kpc, 1.8 and 0.6 respectively and are adopted from \S2.7 page 111 of \cite{2008gady.book.....B}. 
\\ \hline
\end{tabular}
\vspace{0.5cm}
\label{table:models}
\end{table*}

\begin{table*}
\normalsize
\caption[Best-fitting values of the model parameters]
        {Best-fitting values of the model parameters}
 \centering
 \begin{tabular}{ c c  c c }
 \hline\hline 
Galaxy components & Parameter      & Without $v_{\rm LSR}$ prior & With $v_{\rm LSR}$ prior\\ 
                  & (unit)         &                             & \\ \hline
Stellar Halo      & $r_b$ (\kpc)   & $17.5^{+1.2}_{-1.2}$        & $17.2^{+1.2}_{-1.0}$\\
                  & $r_t$ (\kpc)   & $100.4^{+17.7}_{-16.5}$     & $97.7^{+15.6}_{-15.8}$\\ 
                  & $\Delta$ (\kpc)& $8.3^{+7.5}_{-5.6} $        & $7.1^{+7.8}_{-4.8}$\\
                  & $\beta_s$      & $0.6^{+0.2}_{-0.2}$         & $0.4^{+0.2}_{-0.2}$\\                                
Dark Halo         & $c$            & $17.5^{+15.4}_{-7.5}$       & $21.1^{+14.8}_{-8.3}$\\
                  & $\mvir$ ($\times 10^{12} \msun$)             & $0.62^{+0.25}_{-0.21}$   & $0.80^{+0.31}_{-0.16}$\\ 
Disk              & $\mdisk$($\times 10^{11} \msun$)             & $1.5^{+0.6}_{-0.6}$   & $0.95^{+0.24}_{-0.30}$\\
                  & $a$ (\kpc)    & $5.8^{+0.6}_{-0.9}$          & $4.9^{+0.4}_{-0.4}$\\
                  & $b$ (\kpc)    & $0.2^{+0.2}_{-0.2}$          & $0.3^{+0.2}_{-0.2}$\\
Bulge             & $\mbulge$ ($\times 10^{10} \msun$)& $1.2^{+0.4}_{-0.5}$     & $0.91^{+0.31}_{-0.38}$\\
                  & $s$ (\kpc)    & $2.2^{+0.5}_{-0.6}$    & $2.1^{+0.6}_{-0.6}$\\ \hline
Angular velocity  & $\omega_\sun$ ($\kms \kpc^{-1}$)    & $30.2\pm{1.2}$  & -- \\ \hline 
 \hline
 \end{tabular}
 \vspace{0.4cm}
 \label{table:result}
 \end{table*}

\subsection{Density $\rho(r)$}\label{sec:density}
Studies of the morphology of the Milky Way halo suggest that 
a good fit to the stellar halo density distribution is a double power-law
with a shallow slope inside a break radius ($r_b$) and a sharp fall-off outside.
For example analyzing the main-sequence turnoff stars, \cite{2008ApJ...680..295B} conclude that a power-law,
$\rho \propto r^{-\alpha}$, with index of $\alpha=2$ for the inner region
and 4 in the outer region with a break at $r\lesssim20\kpc$ is a reasonable representation.
Similar conclusions were also made by \cite{2009MNRAS.398.1757W} 
in their studies of RR Lyrae stars out to 
$100\kpc$ and by \cite{2011MNRAS.416.2903D} for BHB stars out to $40 \kpc$. 
Apart from the SDSS survey, the study by \cite{2011ApJ...731....4S} 
of the main-sequence turnoff stars 
obtained from Canada-France-Hawaii Telescope Legacy Survey 
suggest slightly shallower fall of 3.8 beyond the break radius $r = 28\kpc$. 
More recently, \cite{2013AJ....146...21S} using variable
stars suggest a broken power law but with much smaller break radius of $\sim16\kpc$.
In an agreement with \cite{2009MNRAS.398.1757W, 2011MNRAS.416.2903D}, here we adopt a double power-law fit to the halo density with 
inner-slope of $2.4$ and the outer-slope of $4.5$ with break at radius $r_b$.
As we will discuss later, the drop seen in the $\rsigma$ profile at $r\approx20\kpc$
seems likely to be a consequence of the break in the density. 
Hence, to further investigate this issue we keep the break radius ($r_b$)
as a free parameter of our model.

The outer-most, $r\gtrsim100\kpc$, region of the halo is diffuse and highly 
structured \citep{2007AJ....134.2236S,2011ApJ...728..106S}.
Only a handful of stars have been observed at such large
distances and so the density profile is largely unknown. 
With a catalog of comparatively larger number of outer-most halo stars,
we investigate as to what does the decline of $\rsigma(r)$ after
$r\gtrsim90\kpc$ tells us about the density profile. 
For this, we investigate a truncated model of the outer halo 
with a truncation radius $r_t$. 
A similar approximation has also been made by \cite{2006MNRAS.369.1688D},
but with a forced sharp truncation at $r=160\kpc$.
However, we soften the truncation by using an exponential functional form 
that has a tunable parameter $\Delta$ that determines the strength
of the fall \citep[see also][]{2012ApJ...750..107S}.
We then let the MCMC likelihood fit determine both the position of $r_t$ and the
scale length of the fall $\Delta$.
The first row in the Table \ref{table:models} shows the form
of the density profile that we adopt.  
Note, we assume the logarithmic slope at $r=r_t$ to be
continuous.  This gives $\epsilon = \frac{r_t}{\Delta} - 4.5$,
and the slope at $r\geqslant r_t$ as $\epsilon - r/\Delta$.

\subsection{Anisotropy $\beta(r)$}\label{sec:anisotropy}
The velocity anisotropy ($\beta$) is the most uncertain quantities that enters the Jeans equation, 
and only recently has $\beta(r)$ been measured directly.
The full phase-space studies of the sub dwarfs in \cite{2009MNRAS.399.1223S} 
and main-sequence stars in \cite{2010AJ....139...59B} suggest that the halo 
within $d<10\kpc$ is radial  with $\beta\approx0.7$.
Recently, K12 studied the line-of-sight velocity of BHB stars and found 
that at similar distance range ($r<12\kpc$) the halo is radial $\beta\sim0.5$.
Moreover, K12 also found that the $\beta(r)$ profile of the stellar halo has features which 
include a tangential dip ($\beta=-1.2$) at $r\sim17\kpc$. 
In the range $19\lesssim r/\kpc \lesssim 25$, K12 find the $\beta$ 
to be consistent with zero given the uncertainty.
Further confirmation of this comes from \cite{2013ApJ...766...24D}
who used the HST proper motion information of the main-sequence
stars to find that at similar distance $r\sim25\kpc$ the halo is isotropic $\beta=0.0^{+0.4}_{-0.2}$.
For $r > 25\kpc$, so far there has been no direct measurement of $\beta(r)$.
Therefore, in this regime we have no choice but to assume some model for $\beta$.
A trend that $\beta(r)$ rises from 0 and attains a positive value at large distances
has been reported (see Figure 13 of K12) for the \lcdm stellar halos of \cite{2005ApJ...635..931B} 
and also for the halo stars in the cosmological simulations of \cite{2007MNRAS.379.1464S}.   
We assume a similar trend here, i.e., at distance 
$r>r_2$, $\beta(r)$ is constant and equal to $\beta_s$ value.
However, in between the maximum observed radius $r=25\kpc$
and the distance $r_2$, $\beta(r)$ is assumed to be linear. 
This is more physical than introducing an abrupt transition like a step-function,
In other words, $r_2$ determines the slope of the $\beta(r)$ profile
joining $\beta_s$ and $\beta|_{r=25\kpc}$.
Here , we assume $r_2$ to be $50\kpc$.
This transition is also consistent with the results from the simulation.
However, in both the above mentioned simulations the transitions cease at much closer 
radius than we assume here. 
Formula shown in the second row of Table \ref{table:models}
summarizes our model for the anisotropy profile.

\subsection{Potential $\Phi(r)$} \label{sec:potentials}
The next quantity required for the $\rsigma$ modeling is the 
potential, $\Phi(r)$.
We essentially construct a model of the Galactic potential with three components:  
an oblate spheroidal bulge, an axisymmetric disk described by \cite{1975PASJ...27..533M} and 
a spherical dark halo described by a \lcdm motivated Navarro-Frank-White \citep[NFW,][]{1996ApJ...462..563N} profile.
The formulas are given in Table \ref{table:models}.

The Galactic dark matter halo  assumed to follow an  
NFW profile, is characterized by two parameters, the virial
mass $\mvir$ and the concentration $c$ ($3^{\rm rd}$ row of Table \ref{table:models}).
Here, the over-density of the dark matter compared to the average matter density, $\delta_{\rm th}$, 
is considered to be 340 \citep{1998ApJ...495...80B}. 
The values for Hubble constant $H_0=70.4 \ \text{kms}^{-1}\text{Mpc}^{-1}$ and the matter density of the universe
$\Omega_{\rm m}=0.3$ are taken from \cite{2011ApJS..192...18K}
\footnote{Recently, the Planck collaboration has revised the value of $H_0$ 
downward to $67.3 \ \text{kms}^{-1}\text{Mpc}^{-1}$ \citep{2013arXiv1303.5076P}.
However, we still use the WMAP7 result from \cite{2011ApJS..192...18K} 
for an easy comparison with previous works.};
in other words, we assume the mean density is 
$\Omega_{\rm m} \delta_{\rm th}$ times the cosmological critical density. 

The Galactic disk is thought to have two major components, i.e., 
a thick and thin disks \citep[e.g.][]{1983MNRAS.202.1025G}. 
Generally, the disk is modelled as exponential 
in a sense that the surface density falls exponentially as a function 
of distance from the centre of the Galaxy R and height from the Galactic mid-plane z. 
Since here we use the spherical Jeans equation we 
consider an analytic and easy to use 3D model of the disk, which is provided by a 
flattened disk of \cite{1975PASJ...27..533M} type.
Its functional form is given in a $2^{\rm nd}$ to the last row of Table \ref{table:models}.
There, $a$ and $b$ are its scale lengths and $M_{\rm disk}$ is the mass.

The Galactic bulge is assumed to be spheroidal. 
This provides a reasonable axisymmetric approximation for bulge seen from COBE/DIRBE near-infrared data.
It is also similar to the axisymmetric approximation of \cite{2002MNRAS.330..591B} 
model considered in \cite{2011MNRAS.414.2446M}.
The last row of Table \ref{table:models} presents the mass density of the 
assumed model for the bulge and is taken from Equation 2.207 a and b in \cite{2008gady.book.....B}.
To compute the force generated by such a spheroidal system we use 
Equations 2.129 a and b of \cite{2008gady.book.....B}.
The values for some of the bulge parameters are kept fixed,
such as,  
oblateness parameter $q_b$ and power-law index $\alpha_b$. 
These were adopted from \S10.2.1 of \cite{BnM98}.
We found that at distance greater than 4 times the scale length $s$, 
the spheroidal bulge contribution to the overall potential 
is similar to that of a point mass.

Both the disk and bulge models are functions of radius and polar angle.
But here we work in a framework of the spherical Jeans equation. 
Hence, we only consider the radial component of the force due to disk and bulge,
i.e., along the basis vector $\widehat{e_r}$, which we average over the spherical shells.
The average force exerted on a unit mass at given radius $r$ 
due to all the three components of the Galaxy is modeled by 
\begin{equation}\label{eqn:accln}
\frac{d\Phi(r)}{dr} = 
\langle \nabla \Phi_{\rm bulge}(R,z).\widehat{e_r} \rangle + 
\langle \nabla \Phi_{\rm disk}(R,z).\widehat{e_r} \rangle + 
\frac{d\Phi_{\rm NFW}(r)}{dr}.
\end{equation}

\begin{figure}
\centering
  \includegraphics[width=0.475\textwidth]{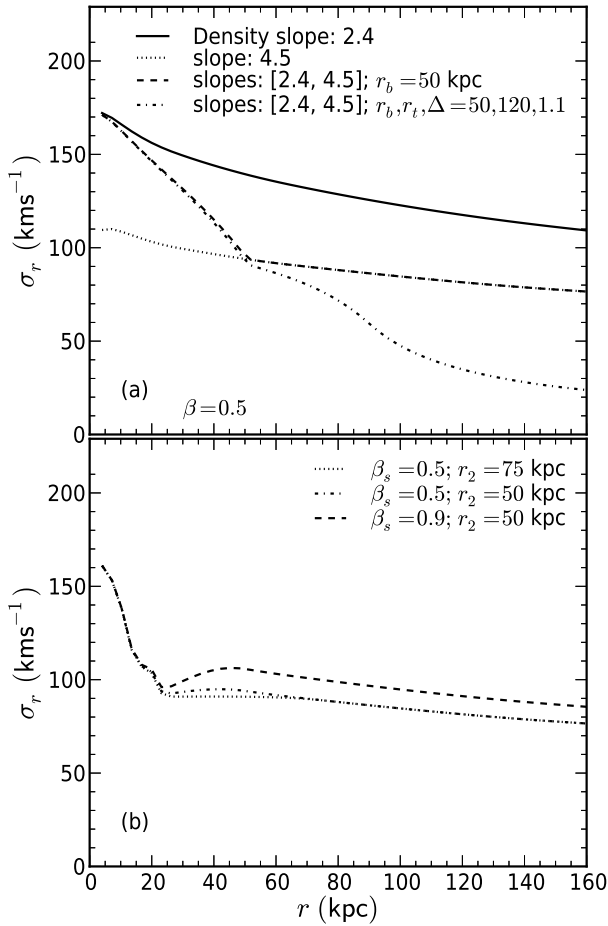} 
    \caption[Properties of $\rsigma$ model]{ Effect of density and anisotropy parameters in the $\rsigma$ model: 
                                           (a) shows an effect of assumed density power-laws 
                                            for the case of constant $\beta$, $c$ and $\mvir$
                                            and (b) shows an effect of assumed $\beta$ profile 
                                            and its parameter $r_2$ and $\beta_s$ for the case of 
                                            double power-law density and constant $c$ and $\mvir$.
                                            In all cases bulge and disk parameters are kept fixed 
                                            as $\mdisk=10^{11}\msun$, $a=4.5\kpc$, $b=0.8\kpc$,
                                            $\mbulge=10^{10}\msun$, $s=1.9\kpc$, $a_b=1\kpc$, $\alpha_b=1.8$ and $q_b=0.6$.}
\label{fig:rsigma_model}
\end{figure}

The dominant contributor to the overall 
potential of the galaxy in the innermost region $ r\lesssim
5 \kpc$ is the bulge, in $5<r<15 \kpc$ is the disk and at
even larger distances is the halo.
Therefore, fixing a potential of any component 
would systematically alter the contribution of the others.
So, we keep the parameters free and allow the data to resolve the degeneracies itself.

Before proceeding with the fit, we first study
the effect of density and anisotropy profiles in the $\rsigma(r)$ model.
The model $\sigma_r(r)$ is obtained by substituting the above described 
density, anisotropy and potential profiles in Equation \ref{eqn:Jeanseq_rsig}. 
In Figure \ref{fig:rsigma_model}a, we demonstrate the role of the assumed density profile
for the case of $\beta=0.5$, $c=10$ and $M_{\mathsf{vir}} = 10^{12} \ M_{\odot}$.
The solid (dotted) line is the case of a single power-law with slopes of 2.4 (4.5) whereas
the dashed line is the case of a double power-law density profile with an inner slope 2.4
and an outer slope 4.5, break being at $r_b=50 \kpc$.
In the figure the $\rsigma$ for the double-power law case attains the values for single 
power-law cases in the inner and outer parts with a sharp transition ceasing exactly at the break radius $r_b$.
Furthermore, the dashed-dotted line in the figure is same as  
in the case of dashed line but with an exponential fall-off beyond $r_t=120 \kpc$ with 
parameter $\Delta=1.1 \kpc$.
Note the $\rsigma$ for the dashed-dotted line and dashed line cases are same
out to $r_b=50 \kpc$ but beyond that the dashed-dotted line declines further 
due to added exponential fall-off in the density profile.
Since the slope in the latter case is a function of $r$ the transition
is smoother than one near the first break $r_b$.
For $r>r_t$, $\rsigma$ decreases gently. 
This suggests that the break in the $\rsigma(r)$ profile is a result of the 
break, $r_b$, in the density distribution.

In Figure \ref{fig:rsigma_model}b, we demonstrate the effect of 
the underlying $\beta$ profile for the case of $c=10$, $M_{\mathsf{vir}} = 10^{12} \ M_{\odot}$, 
density power law index of 2.4 (inner region), 4.5 (outer region) and $r_b=23\kpc$.
The $\beta(r)$ plotted here are taken from second row of Table \ref{table:models}.
The dotted line and dashed-dotted lines show 
$\beta(r)$ with $\beta_s = 0.5$ and $r_2 = 75\kpc$ and $50\kpc$ respectively.
These two runs can be compared to see the effect of $r_2$ on the $\rsigma(r)$ model.
Clearly, the chosen value of $r_2$ that determines the distance at which
$\beta$ saturates and attains constant $\beta_s$ value, is just the measure of the slope of the transition. 
Also, as expected, in the case of $r_2=75 \kpc$ slope is smaller than in the case of 
$r_2=50 \kpc$ as it attains the $\beta_s$ at larger r.
The effect of $r_2$ seems minimal in the overall $\rsigma(r)$ profile.
Hence, we keep it fixed at $50 \kpc$ for the rest of the analysis. 
Furthermore, the above two runs can be compared with the dashed line 
to see the effect of adopted $\beta_s$ which is 0.5 in the former runs and 0.9 in the latter one.
As expected, larger $\beta_s = 0.9$ means the $\rsigma(r)$ shifts up and vice versa.
The $\beta_s$ can systematically shift the $\rsigma(r)$
profile and bias the mass estimate, we keep it free.
 
The model parameters we are interested in measuring 
are the stellar halo density power-law break radius ($r_b$), 
truncation radius ($r_t$), truncation softening ($\Delta$)
and  maximum anisotropy ($\beta_s$); 
dark matter halo concentration ($c$) and the virial mass ($\mvir$);
disk scale lengths ($a$ and $b$) and mass ($M_{\rm disk}$); and 
bulge scale length ($s$) and mass ($M_{\rm bulge}$).
We consider flat priors for all the parameters in the following range:
$r_b\in[8,30] \kpc$, $r_t\in[60,140] \kpc$, $\Delta \in[0,20] \kpc$, $\beta_s\in[-5,1]$,
$c \in [1,60]$, $\mvir \in [0.05,3] \times 10^{12}\msun$, 
$\mdisk \in [0.1,5] \times 10^{11} \msun$,
$a \in [0.1,12] \kpc$, $b \in [0.01,0.5] \kpc$,
$\mbulge \in [0.1,3] \times 10^{10} \msun$, and $s \in [1,3] \kpc$. 

Next, we use the MCMC and likelihood maximization to compute the model parameters of interest. 
The likelihood function we use is 
\begin{eqnarray}\label{eqn:mle_nfw}
L(\theta|\text{Data}) \nonumber = p(\theta|p(\sigma_r|r)) p(\theta|\vterm,l,\sigma_{\vterm}) p(\omega_{\rm LSR}(\theta)),
\end{eqnarray}
where $\theta=\left\{r_b, r_t, \Delta, \beta_s, c, \mvir, M_{\rm
    disk}, a, b, M_{\rm bulge}, s\right\}$ is the set of model
parameters we explore. 
The last term is the prior on $\omega_{\rm LSR}$.
The first term is 
\begin{equation}
 p(\theta|p(\sigma_r|r)) = \prod_{k=1}^{m} p(\rsigma(r_k,\theta)|r_k).
\end{equation}
In this the model $\rsigma(r_k,\theta)$ is given by Equation \ref{eqn:Jeanseq_rsig}. 
The probability $p(\rsigma|r_k)$ is the posterior distribution of $\rsigma$ parameter
at the $k-$th node and is  obtained from Equation \ref{eqn:likelihood} in Section \ref{sec:kinematics}.
In other words, the distribution of each data point in Figure \ref{fig:rsigma}.
This setup avoids assuming gaussian errors for
the measured values, instead we already have the full probability
distribution of $\rsigma$ at each $r$ and we make use of this. 

Our catalog only contains the halo stars and with the halo stars alone 
we cannot construct the rotation curve for the inner region of the Galaxy.
Fortunately, the shape of the rotation curve or the circular velocity $v_{\rm circ}$ in the inner most 
region of the Galaxy $r< R_\sun \kpc$, where the bulge and the disk dominate, 
can be computed from alternative measures such as using tangent point velocities  
\citep[e.g.][]{1995ApJ...448..138M,2007ApJ...671..427M} or the gas rotation curve \citep[e.g.][]{2009PASJ...61..227S}. 
Here we use the terminal velocity curves from 
\cite{1994ApJ...433..687M,1995ApJ...448..138M}, as done e.g. 
in \cite{1998MNRAS.294..429D, 2008ApJ...679.1239W, 2011MNRAS.414.2446M}.

\begin{figure}
 \centering
 \includegraphics[width=0.49\textwidth]{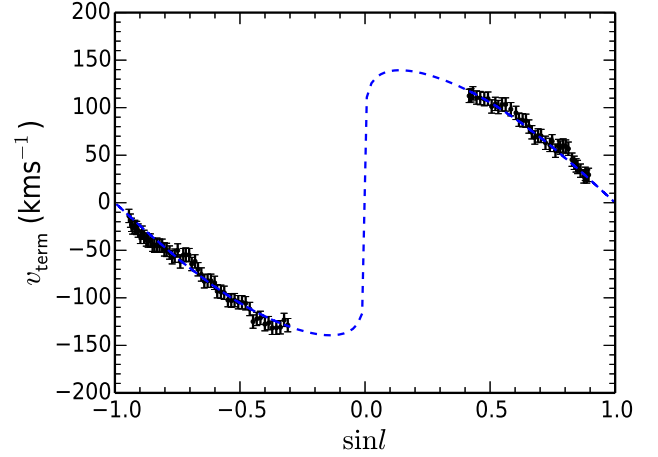}
 \caption[Terminal velocity]{ Terminal velocity $\vterm$ as a function of galactic longitude $l$ : 
          the points are the data taken from \cite{1994ApJ...433..687M,1995ApJ...448..138M}.
          The error bars of $7\kms$ shown are introduced to allow for non-circular motions. 
          The over plotted line is our best fit model resulted from our final MCMC run
          corresponding to the Figure \ref{fig:step_plot}.}
\label{fig:vterminal}
\end{figure}

By measuring the terminal velocity along all lines of sight between Galactic longitudes 
$|l|<\pi/2$ for latitude $b=0^\circ$ it is possible to derive a measurement 
of the rotation curve of the inner Galaxy.
Assuming that the Galaxy is axisymmetric, $\vterm$ as a
function of $l$ is given by 
\begin{equation}
\label{eqn:vterm}
\vterm(l) = \vcirc (R_\sun \sin l ) - v_{\rm LSR} \sin l.
\end{equation}
where,  $v_{\rm LSR}$ = $v_{\rm circ}(R_\sun)$.
Now we define our second term in likelihood as
\begin{equation}
p(\theta|\vterm,l,\sigma_{\vterm}) = \prod_{k} \mathcal{N}(v_{\rm term}(l_k,\theta)|v_{\rm term}^k,\sigma_{\vterm}^k, l^k). 
\end{equation}
There will be effects of non-axisymmetry of the Galaxy and non-circular motion of 
the ISM on the $\vterm^{\rm data}$. 
To take into account this effects, following
\cite{1998MNRAS.294..429D}, we assume 
$\sigma_{\vterm} =  7 \kms$ and  
avoid the region affected by the bar by only using data with  
$|\sin l| > 0.3$ .
The data with assumed uncertainties are shown with black points in Figure \ref{fig:vterminal}

An additional prior we impose is on $v_{\rm LSR}$. 
There is a wide variation in claims about $v_{\rm LSR}$ at $R_\odot$ 
ranging between $184 \kms$ \citep{1998MNRAS.297..943O} to $272 \kms$ \citep{1999ApJ...524L..39M}.
Many of these claims depend on the assumed $R_\odot$ and are normally 
measured using the data within the solar annulus. 
In their studies of masers, \cite{2010MNRAS.402..934M} find that the angular velocity 
is constrained better ranging between $29.9-31.6 \kms \kpc^{-1}$.
As a summary of all these works, we assign a prior with a uniform distribution of
\begin{equation}\label{eqn:prior_wlsr}
p(\omega_{\rm LSR}) = \mathcal{U}(23, 34).
\end{equation}
The range in $\omega_{\rm LSR}$ of $[23,34] \kms \kpc^{-1} $ corresponds to a range in $v_{\rm LSR}$ of 
$[196,289]\kms$ at $R_\odot= 8.5 \kpc$.

\begin{figure}
 \centering
 \includegraphics[width=0.48\textwidth]{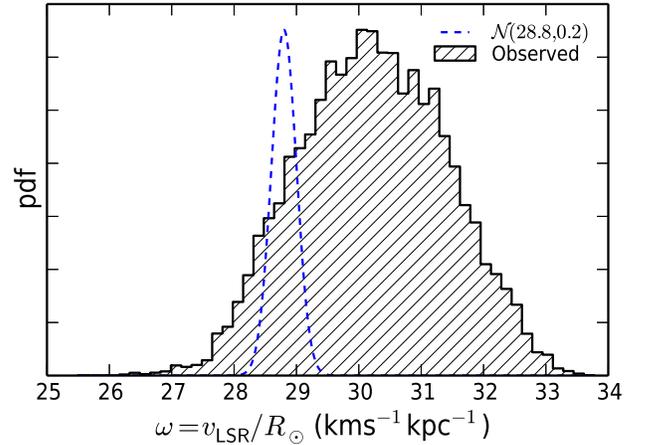}
 \caption[Angular velocity at $R_\odot$]{ Angular velocity ($\omega$) at $R_\odot$: the black histogram
          shows the posterior distribution of $\omega$ obtained from the MCMC run for 
          the case with uniform prior.
          The best fit value of $\omega$ is $30.2\pm{1.2} \kms \kpc^{-1}$.
          Blue dotted line is the normal distribution with a mean value of $28.8 \kms \kpc^{-1}$
          and a dispersion of 0.2 obtained for V$_\odot =12.24\kms$ and $R_\odot = 8.5 \kpc$
          from \cite{2004ApJ...616..872R}, which we later use as a prior.}
\label{fig:omega}
\end{figure}

\begin{figure*}
   \centering
      \includegraphics[width=0.95\textwidth]{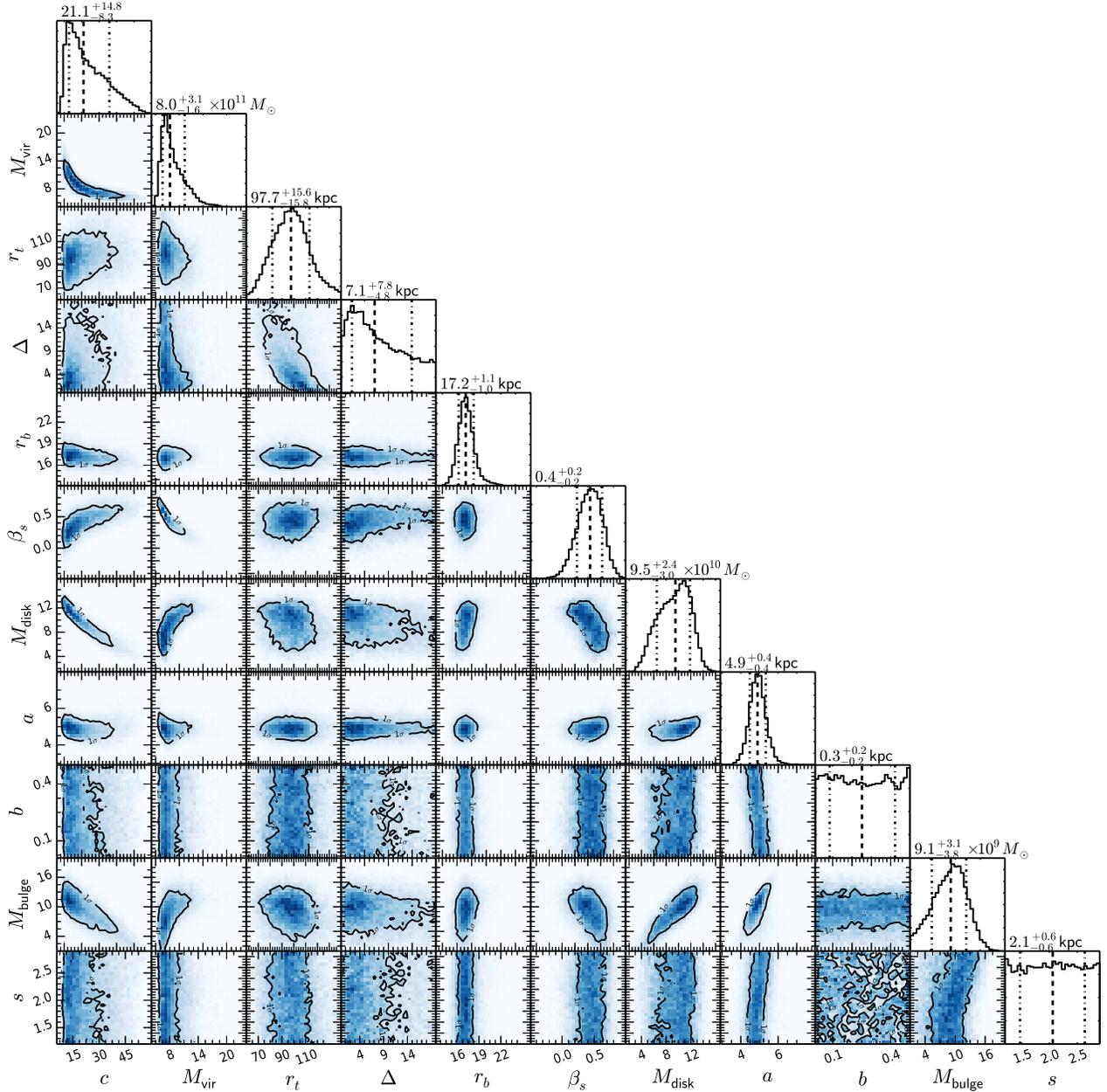}
   \caption[Likelihood maximization of the combined data of the BHB and the giant stars]
    {The joint likelihood and the marginal posterior distributions of the model parameters obtained
    from the MCMC exploration of the combined sample of the halo giant and BHB stars:
    the labels along the horizontal and vertical direction tell the name of the parameters,
    i.e., from the left to the right they are the NFW dark matter halo concentration $c$,
    virial mass $\mvir$ in $10^{11}\msun$, density break radius $r_b$ in \kpc,
    truncation radius $r_t$ in \kpc, truncation softening parameter $\Delta$, anisotropy $\beta_s$,
    disk mass $\mdisk$ in $10^{10}\msun$, disk scale length $a$ in \kpc, disk scale height $b$ in \kpc,
    bulge mass $\mbulge$ in $10^{9}\msun$, and bulge scale length $s$ in \kpc.
    Histograms at the top of each column show the posterior distribution
    of the model parameters named at the bottom of the column whereas the heat maps
    depict the joint likelihood distribution of two parameters named immediately
    below and on the left-most end of the same row.
    Black lines mark the $1\sigma$ confidence contours.
    The values in the title of histograms present the best fit estimates.}
\label{fig:step_plot}
\end{figure*}

\begin{figure}
 \centering
 \includegraphics[width=0.49\textwidth]{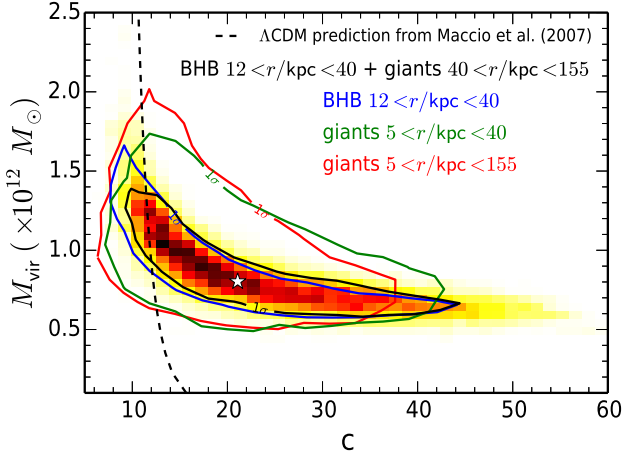}
 \caption[Concentration $\mvir$ at $R_\odot$]
         {Concentration ($c$)--virial mass ($\mvir$) contours for
           $R_{\odot}=8.5 \kpc$:
         red and green contours are for the giants sample, blue contour is for the BHB sample, and
         black contour is for a combined sample of BHB and giant stars 
         in a separate distance ranges as labeled in the figure. Closed lines depict $1\sigma$ region.
         A white star denotes the best fit estimate and pixel plot shows a 
         2D posterior distribution corresponding to the black contour. 
         The black dashed line demonstrates a typical $c-\mvir$ relation 
         predicted by $\Lambda$CDM dark matter simulation. }
\label{fig:c_mvir}
\end{figure}

\begin{figure}
   \centering
   \includegraphics[width=0.48\textwidth]{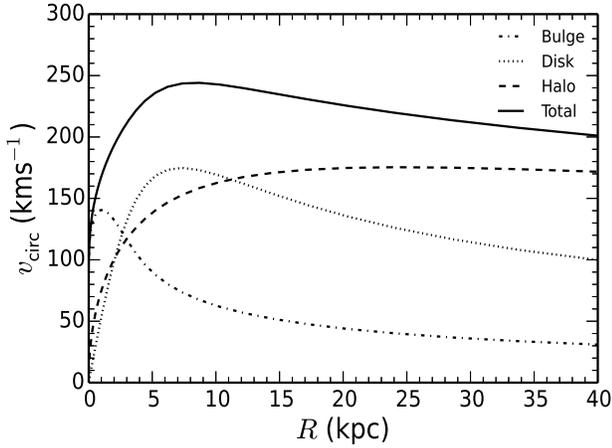}
   \caption[The circular velocity curve of the Milky Way]
             {The circular velocity curve of the Galaxy: 
             the dotted, dashed-dotted and dashed lines are the circular velocity curves
             along the meridional plane, z=0, for the oblate bulge, Miyamoto-Nagai disk
             and NFW halo respectively. The radius $R$ is the distance in the Galactic plane.
             The individual curves are constructed from the best fit estimates of the model parameters.
             The solid line shows the resultant circular velocity curves due to all 
             the three components of the Galaxy.}
\label{fig:vcirc}
\end{figure}

\begin{figure}
  \centering
  \includegraphics[width=0.48\textwidth]{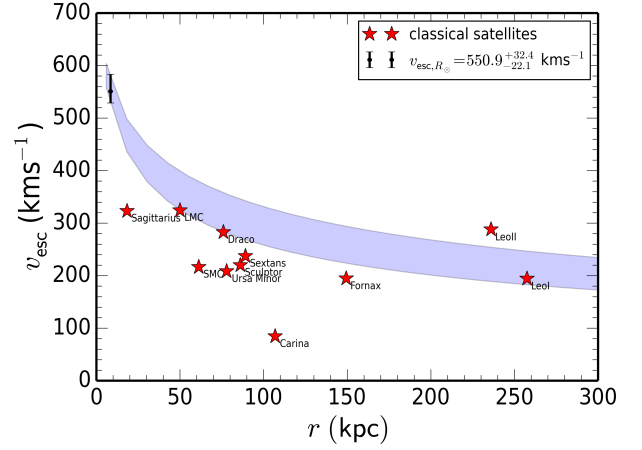}
  \caption[Escape velocity of the Galaxy]
            {Escape velocity of the Galaxy: the observed escape velocity $v_{\rm esc}$ 
             of the Galaxy shown as a function of the galactocentric distance $r$.
             The $v_{\rm esc}$ at $R_\odot$ is shown by a black dot with error bar.
             Red stars shows the total velocities for the named Milky Way classical satellite galaxies
             adopted from Table 1 of \cite{2013MNRAS.435.2116P}.}
\label{fig:vesc_run}
\end{figure}

\begin{figure}
  \centering
  \includegraphics[width=0.48\textwidth]{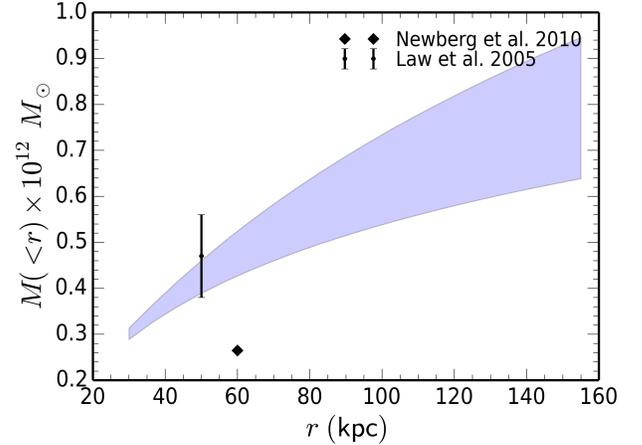}
  \caption[Cumulative mass of the Galaxy]
          {Cumulative mass of the Galaxy: the shade shows the observed mass of the Galaxy $M(<r)$ 
           as a function of the galactocentric distance $r$.
           The black dot with error bar is \cite{2005ApJ...619..807L} estimate of the Galaxy 
           mass within $50\kpc$ obtained by modeling the Sagittarius dwarf spheroidal galaxy tidal streams
           whereas the diamond point is mass within $r<60\kpc$ obtained by \cite{2010ApJ...711...32N} 
           by modeling the Orphan Stream. }
\label{fig:cumu_mass}
\end{figure}

We have the $\rsigma$ run for two different tracers, i.e., giant and BHB stars,
labelled with red and black  in Figure \ref{fig:rsigma}. 
While the $\rsigma$ run of the giant data is measured out to a larger distance $r\sim155\kpc$ 
than of the BHB stars $r\sim 40 \kpc$, the giant stars have comparatively 
larger uncertainties in $r$ and $\rsigma(r)$ than BHB stars. 
Importantly, the $\beta(r\leqslant25\kpc)$ profile that goes into our modeling 
is also unknown for the giant data but known for the BHB data.  
It is therefore more sensible to take the best portion of the data in-hand and combine them.
Therefore, for our final round of measurements, we take an adjoined $\rsigma(r)$:
the BHB data from K12 in the range $12\lesssim r/\kpc \lesssim 40$ and the giants data in the range
$40\lesssim r/\kpc \lesssim 155$.
Note, our model for the $\rsigma(r)$ (see Figure
\ref{fig:rsigma_model}) 
does not predict the flattening-out of the profile in the inner-region $r\lesssim12\kpc$. 
However, the first two $\rsigma(r)$ values in the figure for the BHB sample show a clear flattening.
This is something that needs to be investigated in future, presently we ignore these data points.

During a MCMC run, for every proposed set of values of model parameters, particularly one defining
the potentials, we get a prediction for $\omega_{\rm{LSR}}$ 
(shown in Figure \ref{fig:omega} by a hatched histogram). 
We are able to constrain the $\omega_{\rm{LSR}} =  30.2\pm{1.2} \kms \kpc ^{-1}$.
Interestingly, this is within the uncertainty range of 
\cite{2004ApJ...616..872R} result $28.8 \pm 0.2 \kms \kpc ^{-1}$ for $V_\odot = 12.24 \kms$
obtained using the SgrA$^{\text{\textasteriskcentered}}$'s proper motion.
Also, our estimate falls in the prescribed range $29.9-31.6 \kms \kpc ^{-1}$ in \cite{2010MNRAS.402..934M}, who use the proper motions of masers.
Our measurement therefore can be taken as an independent measurement of the $v_{\text {LSR}}$ at $R_\odot=8.5\kpc$.
Since our constraint of $\omega_{\rm{LSR}}$ has larger uncertainties compare to the one obtained using
SgrA$^{\text{\textasteriskcentered}}$ data, from here on unless otherwise mentioned  
we use $\mathcal{N}( \omega_{\text{LSR}}| 28.8, 0.2)$ as a prior instead of a uniform distribution.

\section{Result and Discussion}\label{sec:result}
Figure \ref{fig:step_plot} displays the marginalized 1D and 2D distributions obtained from the MCMC.  
For a quick referral the table with the best fit estimates is put alongside the figure.
It is also summarized separately in Table \ref{table:result} for both cases, i.e.,
with and without SgrA$^*$ proper motion prior.
The reported best fit values are the medians of the posterior distributions of the 
model parameters whereas 
the uncertainties quoted are computed from $16^{\rm{th}}$ and $84^{\rm{th}}$ percentile values.
The best fit $\rsigma$ model obtained after substituting the above estimates is
shown, against the actual data, by a black dashed line in Figure \ref{fig:rsigma}. 
It can be seen that the best fit represents the data well. 
A small rise in $\rsigma(r)$ data at $r=70\kpc$ seems to be a local effect and
could be due to the presence of some kind of shell like structure at the given distance.
For a proper fitting of such outliers we need to know the underlying $\beta(r)$ run
of the data. 

The Figure \ref{fig:step_plot} nicely demonstrates correlations 
that exist among different parameters we consider here.
For example, one can observe an expected correlation between 
$\beta_s-\mvir$, also known as {\it the mass-anisotropy degeneracy}.
Also, one can see a mild correlation between $\mbulge$ and $\mdisk$;
$\mvir$ and $\mdisk$ etc.
An anticorrelation is seen between $c$, and $\mdisk$ and $\mbulge$.

The best fit estimates of the model parameters enable us to 
construct the rotation curve of the Galaxy (shown in Figure \ref{fig:vcirc}). 
The dashed, dotted, and dashed-dotted lines are the
$\vcirc(R)$ as a function of  
the cylindrical radius $R$ for the halo, disk and bulge respectively 
whereas the solid line is the resultant curve. 
Substituting the best fit rotation curve in Equation \ref{eqn:vterm}, 
gives us the best fit terminal velocity curve.
In Figure \ref{fig:vterminal} it is plotted alongside the
terminal velocity data.

\subsection{Properties of disk and bulge}
In this paper we use a Miyamoto-Nagai disk and a spheroidal bulge. 
The properties of the disk and the bulge are mainly governed by the 
terminal velocity data. They are also quite sensitive to the prior 
on $v_{\rm LSR}$ which in turn depends upon the chosen value of
$R_{\odot}$. Overall the disk mass is around $10^{11} \msun$ and the 
bulge mass is around $10^{10} \msun$. The structural parameters 
like $b$ and $s$ are difficult to measure. 
The bulge mass has a mild dependence on $s$ but other than that 
$b$ and $s$ has little effect on other parameters (see Figure \ref{fig:step_plot}). 
We find that, $a+b$ is well constrained by the data.
This can be seen by the 
strong anti-correlation between them and a narrow spread around it.
Our inability to measure $b$ is due to the following two reasons. 
First, the priors from terminal velocity data and SgrA$^{\text{\textasteriskcentered}}$ we use
basically provide $v_{\text{circ}}(R)|_{z=0}$ and this
sensitive only to the sum $a+b$.  
Second, in our set up the halo kinematics responds to forces averaged in radial shells, 
which decreases sensitivity to $b$.

Since, $b$ and $s$ 
are not well determined by the data, the choice of 
prior becomes important for them. 
Traditionally, a double exponential form is used to fit the disk 
density. By fitting exponential 
disks to mono-abundance populations
\cite{2012ApJ...753..148B} finds scale lengths to
be roughly in range $2-4.5\kpc$ and scale heights to be 
roughly in range $0.2-1 \kpc$. 
Here we use a Miyamoto-Nagai (MN) disk, so to 
facilitate comparison we derive the appropriate scaling factors.  
If we fit the surface density of an MN disk with 
an exponential form, for $2<R/{\rm kpc}<8.5$, we get a scale length 
of about $0.82a$. Fitting the density in vertical direction, in the 
range $0.5<z/ \kpc<2.0$ near Sun, we get a scale height of $1.75b$. 
We adopt a uniform prior for $b$ in range $0-0.5 \kpc$, which is within 
the range expected for exponential disks. The value of $a$ 
that we get is also within the range of expectation.
For $s$ we adopt a uniform prior in range $1-3 \kpc$, which 
is around the value $1.9\kpc$ as suggested by 
\cite{2008gady.book.....B}.

\subsection{Anisotropy in the outer halo}
Our best fit estimate for the anisotropy in outer parts $\beta_s$ is $0.4\pm{0.2}$.  
It is interesting that we are able to constraint $\beta_s$, the reason is as follows.
In the inner most region, the terminal velocity data and prior 
on $\omega_{\rm LSR}$ provide information about the bulge and also to some extent disk parameters. 
In the region $12<r/{\rm kpc}<25$ the BHBs anisotropy is already known and 
the kinematics when put in the Jeans Equation provides estimate for $r_b, \mvir, c$ and disk parameters. 
Now, beyond $r>25\kpc$ where $\beta_s$ is introduced, it is in some sense the only unknown.

\subsection{Mass and concentration of dark matter halo}
We estimate the mass of the dark matter halo to be 
$\mvir=0.80^{+0.31}_{-0.16} \times 10^{12} \msun$ and the 
concentration $c=21.1^{+14.8}_{-8.3}$. 
Corresponding values for the virial radius $R_\text{vir}$ and virial velocity 
are found to be $239.1^{+27.6}_{-16.6}\kpc$ and $120.2^{+13.9}_{-8.3}\kms$ respectively.
It can be seen in Figure \ref{fig:c_mvir}, that there is a strong
anti-correlation between $c$ and $\mvir$. 
The upper bound on $c$ is not as well constrained as the lower bound. 
Simulated virialized halos in \lcdm cosmology, in general predict an inversely proportional 
relation between their mass and concentration 
\cite[e.g.][]{2001MNRAS.321..559B, 2007MNRAS.378...55M, 2008MNRAS.390L..64D, 2011MNRAS.416.2539K}.
The dashed line in Figure \ref{fig:c_mvir} shows one such
relation, $c = 327.3\mvir^{-0.12}$, adopted from \cite{2007MNRAS.378...55M}.   
We see that the prediction of the \lcdm simulation for the dark 
matter halo in range $10^{11}\leqslant\mvir/\msun\leqslant10^{13}$ passes through our measurements.
However, note that the predictions are for pure dark matter simulations, and does not 
include baryonic processes like cooling, star formation and feedback. 
The collapse of gas due to cooling leads to adiabatic contraction of the dark matter halo,
which increases its concentration. Feedback on the other hand can have the reverse effect. 
Therefore, it is difficult to comment if the concentration we get is typical or atypical of the Milky Way sized galaxies.

\subsection{Do the kinematics of the giant and BHB stars result
  in a consistent Galactic potential?}
To answer this we now run the MCMC separately over a subset of our sample of the halo giant and 
the BHB samples taken from Figure \ref{fig:rsigma}.
We select BHB and giant stars in a common radial range, i.e., $r\lesssim 40\kpc$.
For BHB sample we estimate $c = 22.8^{+12.1}_{-7.8}$,  $\mvir = 0.74^{+0.18}_{-0.12} \times 10^{12} \msun$ 
whereas for giant sample we estimate $c = 20.4^{+13.8}_{-8.9}$ and $\mvir = 0.90^{+0.46}_{-0.26} \times 10^{12} \msun$.
The $c-\mvir$ joint-likelihood distributions for the above two data sets  
are shown by the blue and green contours in Figure \ref{fig:c_mvir}, respectively.
The distributions for both the samples seem to be in good agreement,
except for the fact that it is more puffed for giants than BHB sample.
We find that this is mainly due to larger uncertainties in the $\rsigma(r)$
values for the giants in compare to the BHB sample
which we verified by  
swapping the error bars in BHB and giants.
To conclude, the kinematics of the two halo population, namely BHB and giants,
are consistent with 
the fact that they both feel the same Galactic potential,  
at least within the radius $r\lesssim 40\kpc$.

\subsection{Break in slope of stellar halo density profile}
Kinematics of halo stars allow us to constrain the density 
profile of the halo. We had modelled the density profile 
in the inner region by a double power law with fixed slopes, 
but the location of the break radius was kept free.
For BHB we measure $r_b$ to be $17.1^{+1.2}_{-1.0} \kpc$ 
and for giants we measure it to be $22.1^{+4.1}_{-3.1} \kpc$.
A reason for the different $r_b$ for these two halo populations 
could be the uncertainties in the distances, which is larger for the giants than BHBs.
Our estimates for the break radius is slightly smaller than 
$\sim27\kpc$ as claimed in \cite{2011MNRAS.416.2903D} or
$\sim25\kpc$ as found in \cite{2009MNRAS.398.1757W}.
Interestingly, our estimates are in good agreement with 
the recent study of RR Lyrae stars in \cite{2013AJ....146...21S}
who suggest a break in a power law at much smaller radius of $\sim16\kpc$.
A smaller break radius also complies with the study of SDSS main sequence turn-off stars 
in \cite{2008ApJ...680..295B} who conclude that the slope of the density profile 
at $r\lesssim20\kpc$ should be shallower in comparison to the radius outside this range.
Here, a point worth noting is that our estimate of break radius is linked to 
the kinematic features whereas all the above values from the literatures 
are inferred from the studies of the spatial distribution.

\subsection{What more do we learn from the tracers extending out to the ``edge'' of the Galactic halo?}
To find an answer we now run the MCMC hammer over the halo giant stars
spanning $5\lesssim r/{\rm kpc}\lesssim 155$.
The red contour in Figure \ref{fig:c_mvir} shows the corresponding $c-\mvir$ joint distribution,
which is found to almost coincides with the green contours
obtained for a giant catalog with $r\lesssim 40$\kpc. 
Similarly, a comparison of the blue contour in the figure,  
which is for the BHB sample within $12\lesssim r/{\rm
  kpc}\lesssim 40$, with black contour which is for a combined sample of 
BHB and giant stars in distance ranges $12\lesssim r/{\rm
  kpc}\lesssim 155$, show that they are similar. 
This suggests that the giant data between $40<r/{\rm kpc}<155$ 
does not add much to our knowledge of  $c$ and $\mvir$, i.e.,
the potential of the Galaxy. 
This is mainly due to our adoption of a parametrized form 
for the  distribution  of dark mater, namely the NFW
profile. The NFW profile predicts that for $r \gg R_{\rm
  vir}/c$ the falls as $r^{-3}$. Therefore, data that extends out
to about two times the scale radius $R_{\rm  vir}/c$
should be sufficient to constrain the two independent parameters
$M_{\rm vir}$ and $c$ of the NFW profile. 
However, if one wants to really compute the density out to
virial radius, e.g., by non-parametric schemes then
kinematic data till the virial radius would be required.
In our case the distant giant stars ($r>40$ kpc), turn out
to be useful to probe the density distribution of the tracer
population, i.e. the stellar halo.  
We find $r_t = 97.7^{+15.6}_{-15.8}\kpc$ and $\Delta = 7.1^{+7.8}_{-4.8}\kpc$.
It is interesting to note that the hydrodynamical simulations 
\citep[e.g.][]{2006MNRAS.365..747A} to investigate the properties of luminous halos
surrounding isolated galaxies do not predict a truncated halo rather they find that the 
halo extends to the virial radius. In future, our results
regarding the density profile of the outer stellar halo
should be useful for testing theories of stellar halo
formation. Recently, \cite{2014ApJ...787...30D}, using A type
stars from SDSS, find a sudden drop in density profile of
the stellar halo as traced by BHBs and blue stragglers,
lending further support to our kinematics detection of such a drop.  

\subsection{Repercussions of the lighter halo}
The number of sub-haloes of a given mass scales directly with the host halo mass \citep{2008MNRAS.391.1685S}. 
Therefore an accurate estimate of the Galaxy mass has importance in 
understanding \emph{the missing satellite problem}.   
One interpretation of the problem \citep{1993MNRAS.264..201K,1999ApJ...522...82K,1999ApJ...524L..19M,2010arXiv1009.4505B} 
is that the \lcdm paradigm predicts 
larger number of massive subhalos for the Milky Way size halo \citep[e.g.][]{2011MNRAS.415L..40B}. 
The problem can be solved if the mass of the Galaxy is low. 
Figure 5 in \cite{2012MNRAS.424.2715W} allows us to directly compare the host halo mass
against the probability of containing three or less than three 
subhalos with $v_{\rm max} > 30\kms$ (maximum value of the circular velocity).
The relation is inferred from the studies of the halos obtained 
from the Millennium Simulation series, Aquarius and Phoenix projects.
For a direct comparison we scale our measurement of $c,\mvir$ to compute the 
mass $M_{200}$ interior to $r_{200}$ from the center of the halo at which the mean density
is 200 times the critical density.
We obtain $M_{200} = 0.72^{+0.24}_{-0.13} \times 10^{12} \msun $ and 
corresponding concentration $c_{200}= 16.2^{+11.6}_{-6.7}$.
Figure 5 in \cite{2012MNRAS.424.2715W} suggests that 
for the mass equal to our $M_{200}$ there is $\sim70\%$ probability 
that the Galaxy should host three or less than three subhalos with $v_{\rm max} > 30\kms$. 
Interestingly, from observations it is known that there are 
only 3 brightest satellites of the Galaxy namely, Small Magellanic Cloud, 
Large Magellanic Cloud and Sagittarius dwarf have $v_{\rm max} \sim 30\kms$. 
This at least suggests that there is no discrepancy between the observed number 
of luminous satellites with $v_{\rm max} > 30\kms$ and the number predicted by \lcdm.
Furthermore, \cite{2013MNRAS.428.1696V} also concludes that 
for Milky Way mass $\sim 8 \times 10^{11} \msun$, which is
similar to our mass, the number and internal dynamics of 
the dwarf spheroidal satellites of our Galaxy are consistent with the predictions of the $\Lambda$CDM model. 
Therefore, we remark that the scarcity of massive subhalos is not a failure of the
\lcdm paradigm but a repercussion of assuming higher virial mass for the Galaxy.

Another impact of the Galaxy mass is in describing the overall 
dynamics of the orbiting satellite galaxies.
For our low estimate of Galaxy mass, are the satellites still bound is
a natural question to ask.
To study this, we measure the escape velocity $v_{\rm esc}$  using 
\begin{equation} \label{eqn:vesc}
v_{\rm esc}(r) =  \sqrt{2|\Phi(r)|}, 
\end{equation}
where \[\Phi(r) = \langle \Phi_{\rm bulge}(R,z) \rangle + \langle \Phi_{\rm disk}(R,z) \rangle + \Phi_{\rm NFW}(r) .\]
Our estimate of $v_{\rm esc}$ as a function of 
the galactocentric radius $r$ is shown in Figure \ref{fig:vesc_run}.
The stars in the figure show the total velocities for the named Milky Way satellite galaxies.
The velocities are computed from a recent compilation tabulated in Table 1 of \cite{2013MNRAS.435.2116P}. 
Also, the velocities are corrected for our assumption of the velocity of the local standard of rest, 
i.e., $245\kms$ at $R_{\odot}=8.5\kpc$.
Except Leo II, which seems to be marginally unbound. 
We can conclude from the Figure \ref{fig:vesc_run} that all the given 
satellites are bound despite our low estimate for the Galaxy mass.  

In Figure \ref{fig:cumu_mass} we present the cumulative mass $M(<r)$ of the Galaxy.
It is computed using the formula for the centrifugal equilibrium 
\[ M(<r) = \frac{r^2}{G} \frac{d\Phi}{dr}, \]
where $d\Phi/dr$ is taken from Equation \ref{eqn:accln} and uses spherical averaging. 
Infalling satellites are destroyed by their host's gravitational potential resulting in tidal streams.
Attempts to model these streams \citep{2010ApJ...711...32N, 2010ApJ...714..229L, 2012ApJ...744...25C}
also provide an alternate constrain on the Galaxy mass.
In Figure \ref{fig:cumu_mass} the black dot with error bar is \cite{2005ApJ...619..807L}
estimate of the Galaxy mass within $50\kpc$ 
obtained by modeling the Sagittarius dwarf spheroidal galaxy tidal streams.
It is interesting to note that our estimate given the range of uncertainty agrees well this result. 
However, the diamond point which is mass within $r<60\kpc$ obtained by \cite{2010ApJ...711...32N} 
by modeling the Orphan Stream is significantly smaller than our prediction.
One possible reason could be that they model the orbit not the stream and  
the possible misalignment between stream and orbit could bias the result \citep{2013MNRAS.433.1813S}. 

\subsection{Local constraints on mass density, surface density and escape velocity}

The local (at $R_\odot$) dark matter density provides a strong basis for the experimental endeavors for
indirect detection of the dark matter,
see \cite{2013PhR...531....1S} for a review of the topic.
Therefore, determination of the local mass distribution,
the work originally pioneered by \cite{1932BAN.....6..249O},
has recently received a great deal of attention.
Our best fit model of the halo potential allows us to 
compute the local dark matter density $\rho^{\rm DM}_{\odot}$, which we measure to be
$\rho^{\rm DM}_{\odot} = 0.0088^{+0.0024}_{-0.0018} \msun \pc^{-3}$,
equivalent to $0.35^{+0.08}_{-0.07}$ GeV cm$^{-3}$.
Our result is in good agreement with 
the recent estimates of $0.3\pm0.1$ GeV cm$^{-3}$
by \cite{2012ApJ...756...89B}, $0.40\pm0.04$ GeV cm$^{-3}$ by \cite{2011MNRAS.414.2446M}
or $0.389\pm0.025$ GeV cm$^{-3}$ by \cite{2010JCAP...08..004C}.
However, we note slightly lower estimates of  
$0.007\msun \pc^{-3}$ given in \cite{2000MNRAS.313..209H} who utilize a catalog of 
A-F stars obtained from the Hipparcos survey and of $0.0065\pm0.0023 \msun \pc^{-3}$ given in \cite{2013ApJ...772..108Z} who utilize
K dwarf stars from SDSS/SEGUE survey.

Also, we measure the local escape velocity, using the Equation
\ref{eqn:vesc} , to be $v_{\text{esc},R_\odot} = 550.9^{+32.4}_{-22.1}\kms$.
Our estimate seems to be slightly higher but within the range of uncertainties of $544^{+64}_{-46}\kms$ found 
using the high velocity halo stars in \cite{2007MNRAS.379..755S}. 
Moreover, the most recent estimate of $v_{\text{esc},R_\odot}$, again using the high velocity stars, 
is provided to be $544^{+64}_{-46}\kms$ \citep{2014A&A...562A..91P}.
There $v_{\text{esc},R_\odot}$ is defined to be the minimum speed required to reach three virial radius, where
$R_{\text{vir}}=180\kpc$.
For a fair comparison we re-define Equation \ref{eqn:vesc} to be equal to 
$\sqrt{2|\Phi (R) - \Phi (3 R_{\text{vir}} )| } $ and compute 
 $v_{\text{esc},R_\odot} = 528^{+24}_{-17}\kms$ for $R_{\odot}=8.5 \kpc$, which is in the lower range of quoted value in \cite{2014A&A...562A..91P}.

Yet another quantity of interest is whether our disk is maximal 
\citep{1985ApJ...294..494C, 1985ApJ...295..305V}.
A convention \citep{1997ApJ...483..103S, 2013arXiv1309.3276C} is that in a maximal disk 72\% of the total 
rotational support $\vcirc^{\rm total}$ is contributed by a disk $\vcirc^{\rm disk}$, i.e., 
\begin{equation}
F = \left( \frac{\vcirc^{\rm disk}(R_{\rm max})} {\vcirc^{\rm total}(R_{\rm max})} \right) ^2 \gtrsim 0.72,
\end{equation}
where $R_{\rm max}$ is a radius  at which $\vcirc^{\rm disk}(R)$ is maximum. 
For our model, we find $R_{\rm max} = 7.4\kpc$ and $F=0.5$ or 
$\vcirc^{\rm disk}(R_{\rm max})/\vcirc^{\rm total}(R_{\rm max}) =0.7$,
i.e., at this radius $30\%$ of the total rotational support is by a
disk.

Because of a lesser contribution of the disk to the total rotation curve
we find the Galaxy disk at $R_{\rm max}$ to be sub-maximal.
Recently, \cite{2013ApJ...779..115B} using SEGUE dwarfs found slightly higher value of 
$F=0.69$ and concluded the disk to be maximal.
They assumed an exponential disk, for which $F$ is measured at 
$R=2.2R_{\rm d}$ (scale length of the exponential disk).
It is, however, interesting to note that in their studies of 
12 out of 15 distant spiral galaxies, \cite{2005MNRAS.358..503K} find 
on average $F = 0.28$, which is nearly half of our estimate for the Galaxy.

Finally we check if the local (at $R_{\odot}$) 
column/surface density of our best fit model is within an 
expected range. 
The surface density is computed using
\begin{equation}\label{eqn:columndensity}
\Sigma( R ) = \int_{-1.1 \kpc}^{1.1 \kpc} \rho( R, z ) \ dz.  
\end{equation}
The contribution of the disk to the surface density is strongly 
dependent upon our prior for disc scale length $b$, so is not 
really a prediction of our analysis.  
We find $\Sigma^{\rm total}_{\odot}$ is $94.0^{+16.6}_{-20.3} \msun \pc^{-2}$. This is 
slightly higher than $71\pm6 \msun \pc^{-2}$ obtained by 
\cite{1991ApJ...367L...9K} and 
$74\pm6 \msun \pc^{-2}$ obtained by \cite{2004MNRAS.352..440H},
but within the range of large uncertainties we have.
We note that recent measurements by 
\cite{2013ApJ...779..115B, 2013ApJ...772..108Z}
who use the kinematics of the SDSS/SEGUE dwarfs suggest slightly 
lower value of $\Sigma^{\rm total}_{\odot} = 68 \pm4 \msun \pc^{-2}$.

\subsection{Systematics}
The results presented here are potentially subjected to systematic uncertainties that reader should be cautioned.
We use fiducial isochrones for the distance estimation of the giants. 
But there might be systematics associated with them and this will have an effect on our distance estimates.
An independent way to validate our distance measurement would be to use an estimator shown 
in \cite{2012MNRAS.420.1281S}, but this requires proper motions.  
It remains to be seen if any available proper motions of the distant giants 
are accurate enough for this method to be successfully applicable. 
Additionally, while measuring the distance to the halo K-giants we do not take into account 
uncertainties in the reddening estimate. 
We simply use colors dereddened according to \cite{1998ApJ...500..525S} extinction maps.
This could introduce systematics in our distance estimate.
However, we believe that such systematics, if any, would be insignificant. 
Firstly, the \cite{1998ApJ...500..525S} maps are accurate for high latitude stars and $98\%$ of our sample has $|b|>20^{\circ}$. 
Secondly, although, \cite{1998ApJ...500..525S} maps provide total extinction, 
they are appropriate for halo stars as most of the dust is confined to the disk.  

We use a sample of giants from SEGUE, which are subjected to a proper motion restriction of 11 mas/year \citep{2009AJ....137.4377Y}.
This can potentially introduce a systematic bias for nearby giants, but the giants that we use for our main analysis 
are beyond $40 \kpc$ and for them the above mentioned proper motion limit safely includes all bound halo stars. 

We assume that each tracer population is in Jeans equilibrium.
In any case, it is important to note that we can determine the Galactic potential only to the extent that 
the phase space distribution of tracer stars is in equilibrium \citep{2013NewAR..57...29B}. 
However, if the tracer population under study is a superposition of multiple populations, then the Jeans Equation should be applied separately for each population.  
In earlier studies with BHB stars \citep{2013MNRAS.430.2973K,2013ApJ...763L..17H}, correlation between metallicity and kinematics of halo targets was found. 
A similar correlation could exist for giants. 
So ideally, if one has a larger sample of stars one should treat metal rich and poor populations separately. 
Morever, close to the disk the potential is not spherically symmetric, so strictly speaking the spherical 
Jeans Equation is not valid. This can potentially bias the estimate mass. 
This should be investigated in future.

We model the disk by the Miyamoto-Nagai form.
In reality, the disk is described better by double exponential functions. 
Moreover, the disk of the Milky Way is probably a superposition of multiple populations each with different scale height and length. 
These facts  can potentially bias our results.
Finally, throughout our analysis we assume $R_{\odot}=8.5$ and systematics due to our adopted value can be expected.
Claims for a wide range in $R_{\odot}$ between $\sim7.7-8.8 \kpc$
exist in the literatures \citep[e.g.,][etc]{1993ARA&A..31..345R, 2004ASPC..316..199N, 2013IAUS..289...29G, 2013arXiv1309.2629F, 2014ApJ...783..130R}. 
To study an influence of our adopted value of $R_{\odot}$ we reanalyze our data with $R_\odot = 8 \kpc$.
Firstly, we measure the $\rsigma(r)$ run, which we find to be similar to one shown in Figure \ref{fig:rsigma}.
Secondly, we run the MCMC model fitting to the above $\rsigma(r)$ run. 
Assuming $R_\odot = 8 \kpc$ means the prior in $v_\text{LSR}$ changes 
and this can have a significant influence in our measurement. 
It is mainly because we do not have halo data within $r\lesssim12\kpc$ and the information 
about disk and bulge properties mostly comes from the assumption about  $v_\text{LSR}$ at $R_\odot$.
We find that for $R_\odot = 8 \kpc $ our model parameters 
$\mvir = 1.2\pm{0.3} \times 10^{12} \msun$, $c = 15.2^{+5.6}_{-3.2}$,
$\mdisk = 0.71^{+0.11}_{-0.13} \times 10^{11} \msun$, $a = 3.0\pm{0.6} \kpc$,
$\mbulge = 0.70^{+0.31}_{-.32} \times 10^{10} \msun$ and $s = 2.1\pm{0.6} \kpc$ whereas 
$r_b, r_t, \Delta$ and $\beta_s$ remain unchanged.
   
Recall that our final results shown in Table \ref{table:result} or Figure \ref{fig:step_plot} are obtained for a combined sample 
of two halo populations, i.e., BHB within a radius $r\lesssim40\kpc$ and giants outside $r\gtrsim 40\kpc$.
As discussed earlier in Section \ref{sec:density} 
the density profile for same sample of BHB stars as ours is already computed in \cite{2011MNRAS.416.2903D}
and this is what we assume in our analysis. 
In the region where giants data is used we assume density profile to be 
a power law with slope of -4.5, which is consistent with the findings for BHB \citep{2011MNRAS.416.2903D} 
and RR Lyrae stars \citep{2009MNRAS.398.1757W}. 
However, there is no direct measurement of the density of halo giants.
So, to test the sensitivity of our results to density profile, we run 
the MCMC simulation for two different values of slope indices, -4 and -5. 
From our fits we find that $\mvir$ and $c$ are directly proportional to the density slope. 
For each linear step of -0.5 from -4 to -5 we find that $\mvir$ increases by $\sim30\%$ for each step whereas 
$c$ increases by $\sim20\%$. 
For the same steps from -4 to -5, we find that $\mdisk$ and $\mbulge$ decreases by $\sim22\%$ and $\sim20\%$ respectively.
Quantities such as $r_b$, $r_t$, $\Delta$ and disk and bulge scale parameters remain unchanged. 
Moreover, shape of the posterior distributions of the parameters and hence uncertainties remain same for all the cases.
It should be noted that the value of $\beta_s$ remains nearly the same for both the cases, 
i.e., $0.5\pm{0.2}$ for -4 and $0.4\pm{0.2}$ for -5. 
This is not unexpected, as data in $12<r/\kpc<25$ 
for which observed $\beta$ is available is enough to constraint 
$M_{vir}$ and $c$.  So any change in density profile affects the
enclosed mass. When data beyond $25\kpc$ is added the increase
in power-law index is compensated by the increase in mass so as
to leave  $\beta_s$ unchanged.

Before concluding, it is worth mentioning that we noted 
the parallel work by \cite{2014ApJ...785...63B}.
It has some similarities with our works, e.g.,
uses of the Jeans Equation, SDSS/SEGUE BHB and giant stars catalog etc.     
While estimating the mass using the Jeans Equation, 
we note that they use the density profile of the spectroscopic sample which has a selection bias.  
In reality, the Jeans Equation requires underlying density profile of the tracers and not of the sample. 
Lastly, it has been found in the simulations that halo stars, satellites and the 
dark matter halo have different orbital properties \citep{2006MNRAS.365..747A, 2007MNRAS.379.1464S}. 
Hence, assuming a constant anisotropy for both field stars and satellites could introduce 
additional systematic uncertainties in their mass estimate.
However, it should be noted that they do not model the disk, bulge and halo separately,
but only provide an estimate for the total mass $M(<r)$ within some radius.  

\section{Conclusion} \label{sec:conclusion}
A spectroscopic survey such as the SEGUE provides us with a
large catalog of distant and different tracer populations. 
Here, we complemented the BHB star catalog \citep{2011ApJ...738...79X} 
of the halo tracers with a catalog of K-giant stars. 
The position and line-of-sight velocities of these tracer populations,
the terminal velocity curve and the proper motion of the SgrA$^{\text{\textasteriskcentered}}$ 
allow us to constrain the mass model and tracer properties
of the Galaxy. 
This also allows us to break the degeneracy due to 
the varying relative contribution of the bulge-disk-halo to the rotation curve with the distance.  
Our presented estimates are the marginalized results over all possible values of anisotropy parameter,
thus, also takes  {\it the mass-anisotropy degeneracy} into account.

Our main results considering solar motion with respect to the local standard of rest as 
$U_\odot =+11.1 \kms$, $V_\odot =+12.24 \kms$, $W_\odot =+7.25 \kms$ and its position 
from the centre of the Galaxy at $R_\odot = 8.5 \kpc$, 
are summarized in Table \ref{table:result} and discussed in Section \ref{sec:result}.
Following paragraphs highlights some of our main findings.

\paragraph{Stellar halo} The kinematics of the halo stars
enables us to model the density profile of the stellar halo. 
We model the halo number density to be a double power law with inner slope of -2.4
and outer slope of -4.5 with a break occurring at radius $r_b$. 
We find $r_b = 17.2^{+1.1}_{-1.0}\kpc$.
The break in the radial velocity dispersion profile is found to correspond 
to the break in the density. 
The mass estimate is found to be sensitive to the break radius $r_b$. 
The giant data reveals that the outermost halo stars 
have a small velocity dispersion but interestingly this suggests a truncation of the stellar halo 
density rather than a small overall mass of the Galaxy.
We find that the stellar halo has an exponential truncation 
that starts at radius $r_t = 97.7^{+15.6}_{-15.8}\kpc$ and
has a scale length of $\Delta = 7.1^{+7.8}_{-4.8}\kpc$.  
Direct estimation of the density profile using photometry,
also seems to support the features in the density profile of
the halo that we see using kinematics. For example,
\cite{2013AJ....146...21S} using RR Lyrae report a break at
$r_b=16 \kpc$ and \cite{2014ApJ...787...30D}
using A type stars suggests a sharp fall beyond 50 kpc.
Finally, our modelling also enables us to place some limits
on the anisotropy in the outer halo and we find it to be 
$\beta=0.4^{+0.2}_{-0.2}$.  
  
\paragraph{Dark matter halo} We find that the mass of the 
dark matter halo is $0.80^{+0.31}_{-0.16}\times 10^{12} \msun$,
and concentration is $21.1^{+14.8}_{-8.3}$. 
The upper uncertainty on concentration was found to be quite large.  
The mass estimate is lower and concentration estimate is higher than 
what has been previously measured. For a lower mass like ours, 
recent studies by \cite{2012MNRAS.424.2715W} and \cite{2013MNRAS.428.1696V} 
suggest that the number of massive satellite galaxies, 
i.e., with $v_{\rm max}>30\kms$, observed in the Galaxy matches predictions  
for \lcdm halos of similar mass, potentially solving
\emph{the missing-satellite problem} at the high mass end. 
We also discussed other repercussions of the more
concentrated and lighter Galaxy,
 e.g., all the classical satellite galaxies within 
the Galaxy were found to be bound. 

\paragraph{Disk, bulge and local parameters} 
Additional data in the inner region, i.e., the
gas terminal velocity curve taken from \cite{1994ApJ...433..687M,1995ApJ...448..138M} 
and the proper motion of the SgrA$^{\text{\textasteriskcentered}}$
taken from \cite{2004ApJ...616..872R} also 
enable us to constrain the bulge and disk properties. 
The disk assumed to be of Miyamoto-Nagai form has a mass 
of $0.95^{+0.24}_{-0.30} \times10^{11}\msun$ 
and a scale length of $4.9^{+0.4}_{-0.4} \kpc$, while  
the bulge has a mass of $0.91^{+0.31}_{-0.38} \times10^{10}\msun$.

Furthermore, it is important for a mass model of the Galaxy
to agree with standard local constraints such as the escape velocity, total column density integrated over $|z|\leqslant1.1\kpc$ and the dark matter density.
The escape velocity and the local dark matter density
are in agreement with recent claims in the literature. 
Recent estimates of column density are slightly lower but within our quoted range.
If the SgrA$^{\text{\textasteriskcentered}}$ constraint is not used, our analysis
independently suggests the angular velocity at the Sun to be 
$\omega_{\rm{LSR}} = 30.2\pm{1.2} \kms \kpc^{-1}$. 

In the end, we reiterate that our estimates of the mass parameters
sensitively depend on the choices of $R_\odot$ and the outer-power law index of the tracer number density.
The systematic uncertainties are of the order (e.g. $\mvir$) 
and sometimes larger (e.g. $a$) than the random uncertainties. 
For example, we find that $\mvir$ and $c$ are directly proportional to the density slope. 
For each linear step of -0.5 from -4 to -5 we find that $\mvir$ increases by 
$\sim30\%$ for each step whereas $c$ increases by $\sim20\%$ whereas  
$\mdisk$ and $\mbulge$ decrease by $\sim22\%$ and $\sim20\%$ respectively.
Further systematics inherent to the choices of parameters and assumptions we make 
in our anaylsis are discussed in detail in \S 5.9.

\section*{Acknowledgement}
PRK acknowledges the University of Sydney International Scholarship and ARC grant DP140100395 for the financial support.
GFL acknowledges support for his ARC Future Fellowship (FT100100268) and through the award of an ARC grant DP110100678.
SS and JBH are funded through ARC grant DP120104562 and ARC Federation Fellowship.
Also, a sincere thanks to \cite{Hunter:2007} and \cite{PER-GRA:2007}
for their brilliant softwares which were used extensively in this article.
We sincerely thanks the referee and Pascal Elahi for constructive comments.

\appendix
\section{Diagnostic criteria used for selecting the giant stars} \label{sec:diagnostic}
\begin{figure}
  \centering
  \includegraphics[width=0.48\textwidth]{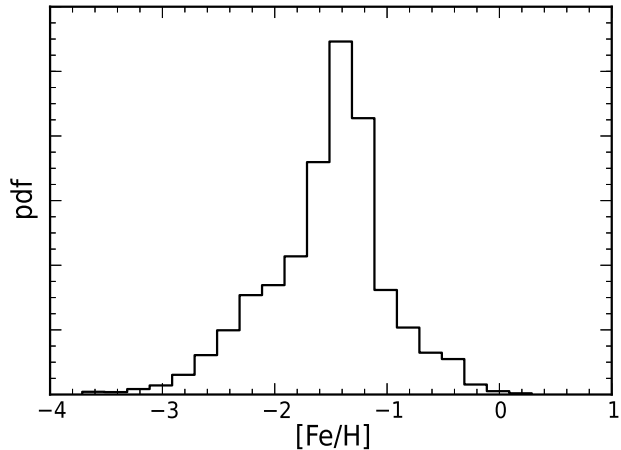} 
    \caption[Metallicity distribution of K-giant stars]
            {Metallicity (\feh) distribution of the catalog of giant stars.}
\label{fig:feh_dist}
\end{figure}
Out of 5330 candidate giant stars, the SSPP classifies 22 as K-giant, 223 as red K-giant,
3111 as l-color K-giant, 536 as proper-motion K-giant and 1438 as M-giant. 
For reference, the \feh distribution of 5330 stars is shown in 
Figure \ref{fig:feh_dist}, which is also used as a metallicity prior while distance measurement.
\begin{figure}
  \centering
  \includegraphics[width=0.480\textwidth]{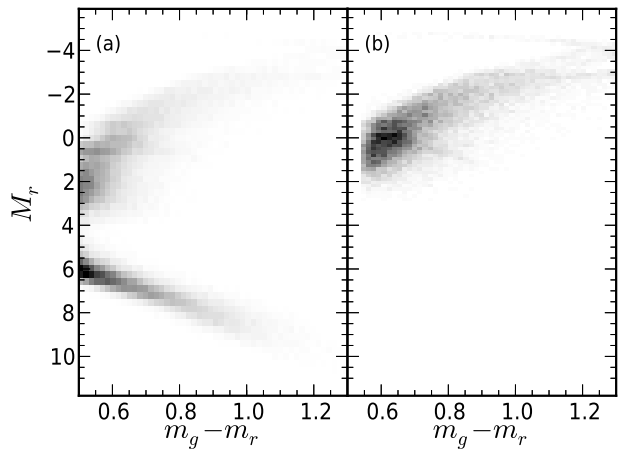} 
  \caption{Color-magnitude diagram of the GALAXIA data: 
           (a) CMD of total sample
           (b) CMD after selecting stars according to Equation \ref{eqn:setA}.}
\label{fig:cmd_galaxia}
\end{figure}

To know how well the criteria listed in Equation \ref{eqn:setA} cleans our catalog 
we use code GALAXIA \citep{2011ApJ...730....3S}.
Galaxia, has an analytical model based on
\citet{2003A&A...409..523R} and uses isochrones from the 
Padova database \citep{2008A&A...482..883M,1994A&AS..106..275B} 
to compute photometric magnitudes for the model stars.
First, we generate stars over an area of 8000 square degree toward the North Galactic Pole.
For a fair comparison with observed data, the mock sample is then convolved with 
the typical errors in the photometric and stellar properties quoted by SDSS/SEGUE. 
According to the provided specifications, the uncertainties in 
the SSPP stellar parameters: the effective temperature ($T_{\rm eff}$), 
the surface gravity ($\log g$) and the metallicity ([Fe/H])
are respectively 117 K, 0.26 dex and 0.22 dex, which in reality also 
depend on the type and the signal-to-noise ratio of the spectra.
Here, we do not deal with the spectra and thus, just assign gaussian random 
error to the stellar parameters and the metallicity with dispersion 
chosen to be same as above uncertainty values. 
We also convolve the mock data with the error in magnitude given by 
$\Delta m = 0.015 + 10^{-3 + 0.4(m - 22.6)}$.
This relation roughly matches to the errors in SDSS photometry 
\citep[Figure 7 with a systematic of 0.015]{2008ApJ...673..864J}.

The above ``ideal'' data once convolved with the observational errors are shown in 
a color-magnitude diagram (CMD) in Figure \ref{fig:cmd_galaxia}(a).
Figure \ref{fig:cmd_galaxia}(b) is again CMD for the mock data but after  
imposing the set of cuts given in Equation \ref{eqn:setA} that we apply to obtain SEGUE giants. 
The final sample of stars remained in our mock catalog is found to 
contain negligible amount $<0.5\%$ of dwarfs contamination.
This assures that our selection criteria performs well, at-least for the theoretical data.

\section{Distance Estimation}\label{sec:distance_appendix}

\begin{figure*}
  \centering
     \includegraphics[width=0.485\textwidth]{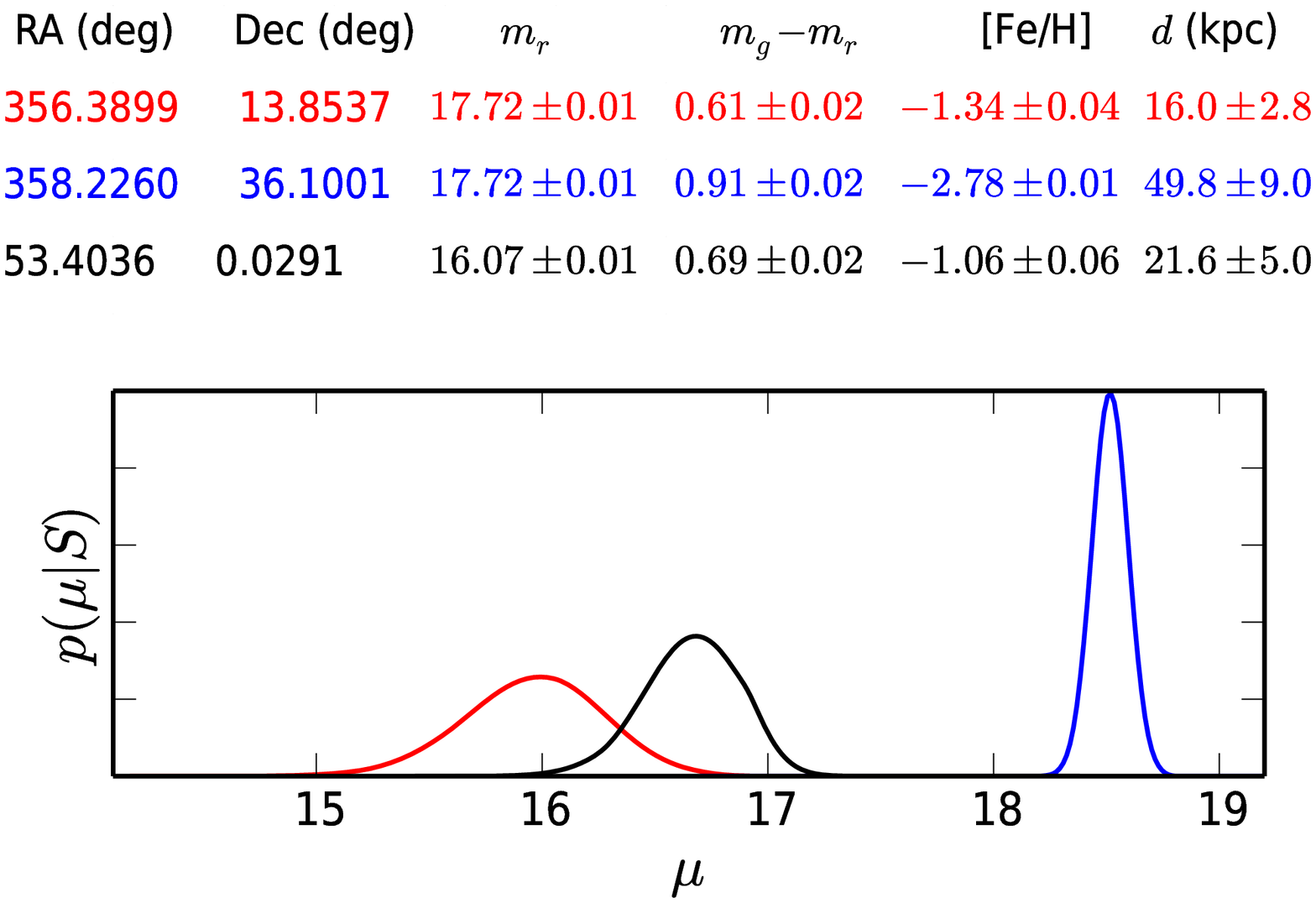}
     \caption[Distance probabilities of sub-sample of giants ]
           {Likelihood distributions of the distance moduli of 3 giant stars and the effect of priors: 
            the red, black and blue colored lines display the posteriors of the distance probabilities 
            $p(\mu|S)$ for 3 different stars with coordinates Right Ascension (RA) and Declination (Dec) given in J2000 epoch;
            magnitude $r$; color $g-r$; metallicity [Fe/H] in dex; and distance $d$ in \kpc
            shown in colored texts above the plots. }
 \label{fig:muposterior}
\end{figure*}
The procedure for distance estimation is same as in
\citet{2014ApJ...784..170X} but was implemented independently.
We rely on the Bayes rule. It allows us to update an initial probability 
(prior) into a revised probability (posterior) 
and is written for our case as  
\begin{equation}\label{eqn:bayes}
p(\mu | S ) \propto p( S| \mu) \times p(\mu).
\end{equation}
Here, $ \mu = \text{apparent magnitude} \ (m) - \text{absolute magnitude} \ (M) $
is a distance modulus, whereas $S$ represents a set of 
observables given by, $S = \{m,c,\feh \}.$
The color c is assumed to be a function of $M$ and \feh. 
In Equation \ref{eqn:bayes}, the posterior $p(\mu | S )$ is the 
probability distribution of $\mu$ given the data $S$; 
the prior $p(\mu)$ is the information about $\mu$ known a priori; 
and the likelihood function $p( S| \mu)$ gives the likelihood of obtaining 
the data $S$ given the $\mu$.
Here, the likelihood function can be more explicitly written as
\begin{equation}\label{eqn:mu_likelihood}
p(S|\mu) = \iint \mathcal{P}(S|\mu) \  p(M) \ p(\feh) \ dM \ d\feh.
\end{equation}
where the functional form for the probability $\mathcal{P}( S | \mu)$ is given by
\begin{equation}
\mathcal{P}(S|\mu) = \mathcal{P}(m, c,\feh |\mu) = \mathcal{N}(c| c',\sigma_{c}) \times \mathcal{N}(\mu+M|m',\sigma_{m}) \times \mathcal{N}(\feh|\feh',\sigma_{\feh}).
\end{equation}
Here $m',c',\feh'$ are the observables for each star 
and $\sigma_m$, $\sigma_c$ and $\sigma_\feh$ are associated uncertainties respectively. 
In Equation \ref{eqn:mu_likelihood}, the probability function $\mathcal{P}( S | \mu)$ are
weighed with the luminosity $p(M)$ and metallicity $p(\feh)$ prior probabilities.
The theoretical \citep{2002PASP..114..375S} and observational evidence 
\citep{2000ApJ...529..936L} suggests that the luminosity function of the giants follow a power law. 
Therefore, fitting a power-law to the luminosity function of the RGB stars in the globular clusters, namely M5 and M30, 
shown in the figures 2 (for M5) and 4 (for M30) of \cite{2000ApJ...529..936L} 
we determine a common slope of 0.32. 
This leads to the final expression for a prior on the luminosity function given by
$p(M) = 10^{0.32M}/17.79$, which is normalized to unity in the data range $M\in[-3.5,3.5]$.
The magnitude of the RGB star has been found to be independent of 
the metallicity content \citep{2002PASP..114..375S} and 
hence we neglect the effect of the stellar metal content in our luminosity priors.
A prior for the metallicity $p(\feh)$ is chosen to be same 
as the \feh distribution of the data, shown in Figure \ref{fig:feh_dist}.
The color $c$ is a function of magnitude $m$ and metallicity [Fe/H].
Hence, we do not need to explicitly assume a prior for the color but 
a relation between $c$, $m$ and $\feh$ has to be defined. 
We derive this relation from the available isochrones of 3 globular clusters, 
namely M92, M13, M71 and an open cluster NGC 6791 
taken from \citet[][and the references therein]{2008ApJS..179..326A}. 
The [Fe/H] values for M92, M13 and M71 are taken to be
-2.38, -1.60 and -0.81 \citep{2003PASP..115..143K, 2008ApJS..179..326A} respectively 
whereas for NGC6791 it is assumed to be +0.40 \citep{2008ApJS..179..326A}.
The distance moduli are taken to be 14.64 for M92, 14.38 for M13 \citep{2000ApJ...533..215C}, 
12.86 for M71 \citep{2002A&A...395..481G} and 13.02 for NGC6791 \citep{1996AJ....112.1487H}.
The color-magnitude-metallicity relation hence obtained are 
similar to Figure 2 of \cite{2013PASA...30....8K} and Figure 5 of \cite{2014ApJ...784..170X}.
The obtained fiducials (color-magnitude curves) are 
are found to be well approximated by the $7^{\text{th}}$ order polynomial fit.
The coefficients of the polynomial are then linearly interpolated  
in order to fill the gaps between the available isochrones of the clusters.
Given a color $g-r$ and [Fe/H] for a star the interpolated fiducial sequences
are then used to compute the corresponding value of the magnitude $M_r$.

Finally, to compute the posterior distribution $p(\mu|S)$ for an individual star (Equation \ref{eqn:bayes})
we also need to consider a distance prior $p(\mu)$. 
Recent observational evidences \cite[e.g.][etc]{2009MNRAS.398.1757W, 2011MNRAS.416.2903D, 2012ApJ...756...23A}
support a broken power-law for the density distribution of the halo stars.
As a convenient summary of all these works, we assume
$\rho \propto r^{-\alpha}$, with inner slope of 2.4, 
outer slope of 4.5 and break at radius $25\kpc$. 
The change of variables is done using, $p(\mu)d\mu = p(r)dr = 4\pi r^2 \rho(r) dr.$  
Using the photometric parallax relation $d/\text{kpc}=10^{(\mu/5 - 2)}$
and assuming that $d\approx r$ our final expression for the distance prior is  
$p(\mu) = \frac{4}{5}\pi \ln(10) r^3 \rho(r).$
\begin{figure}
    \centering
    \includegraphics[width=0.48\textwidth]{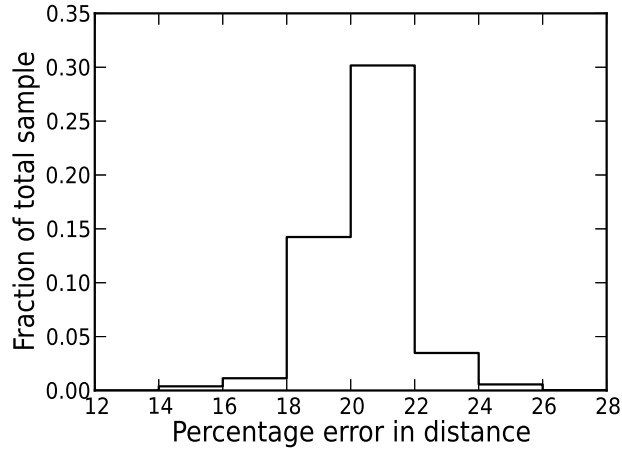} 
        \caption[Errors in distance estimation of K-giant stars]
            {Percentage errors in distance estimation of K-giant catalog.}
\label{fig:dist_err}
\end{figure}
The uncertainties in distance measurement of our catalog are 
shown in Figure \ref{fig:dist_err}. 
This is computed from the $16^{\rm th}$ and $84^{\rm th}$ percentiles of $p(\mu|S)$ of stars.

As an example of this approach, in Figure \ref{fig:muposterior} 
we present results for 3 arbitrary giants from our catalog. 
The colored texts at the top of the figure provide the position
in Right Ascension (RA) and Declination (Dec), magnitude $m_r$, color $m_g-m_r$,
metallicity [Fe/H] and distance $d$ for the corresponding stars shown in the same color on the 
immediate figures underneath, which show the posteriors of distance probabilities $p(\mu|S)$.

\vspace{0.5cm}

\bibliographystyle{apj}
\bibliography{ms}
\end{document}